\documentclass[smallcondensed]{svjour3}
\usepackage[fleqn]{amsmath} 
\usepackage{graphicx}    
\usepackage[misc]{ifsym} 
\usepackage[bookmarksopen,bookmarksdepth=2,breaklinks=true]{hyperref}
\usepackage{breakurl}
\usepackage{bm}          
\usepackage{tabularx}    
\usepackage{natbib}
\bibpunct[,]{(}{)}{;}{a}{}{,}

\newcommand{\rhatEM}{\bm{\hat r}_\mathrm{EM}}
\newcommand{\phat}{\bm{\hat p}}
\newcommand{\Nmax}{n_\textrm{max}}
\newcommand{\OCnm}{\bar{C}_\mathrm{nm}}
\newcommand{\OSnm}{\bar{S}_\mathrm{nm}}
\newcommand{\Cnm}{C_\mathrm{nm}}
\newcommand{\Snm}{S_\mathrm{nm}}
\newcommand{\Vnm}{\bar{V}_\mathrm{nm}}
\newcommand{\Nnm}{N_\mathrm{nm}}
\newcommand{\taumO}{\tau_\mathrm{mO}}
\newcommand{\taumR}{\tau_\mathrm{mR}}
\newcommand{\Ncmb}{{\bm N}_\mathrm{cmb}}
\newcommand{\LBCRS}{{\bm l}_\mathrm{BCRS}}
\newcommand{\LLCRS}{{\bm l}_\mathrm{LCRS}}
\newcommand{\LPA}{{\bm l}_\mathrm{PA}}
\newcommand{\SBCRS}{{\bm s}_\mathrm{BCRS}}
\newcommand{\SGCRS}{{\bm s}_\mathrm{GCRS}}
\newcommand{\STRS}{{\bm s}_\mathrm{TRS}}
\newcommand{\Dgrav}{\Delta_\mathrm{grav}}
\newcommand{\Datm}{\Delta_\mathrm{atm}}

\newcommand{\arcsec}{{{}^{\prime\prime}}}
\newcommand{\Deg}{{}^{\circ}}
\newcommand{\rMAcdotl}{\left(\bm{\hat{r}}_\mathrm{MA}\cdot\bm{\hat l}\right)}

\journalname{Celestial Mechanics and Dynamical Astronomy}
\date{Received: 14 Jan 2016 / Accepted: 25 Jun 2016 \\  \\
The final publication is available at Springer via\\ \url{http://dx.doi.org/10.1007/s10569-016-9712-1}}

\begin{document}
\title{Determining parameters of Moon's orbital and rotational motion
  from LLR observations using GRAIL and IERS-recommended models}
\titlerunning{Determining parameters of Moon's orbital and rotational motion}

\author{Dmitry A. Pavlov  \and
        James G. Williams \and
        Vladimir V. Suvorkin}

\authorrunning{D.\,A. Pavlov, J.\,G. Williams, V.\,V. Suvorkin}

\institute{D.\,A. Pavlov ({\Letter}), V.\,V. Suvorkin \at Institute of Applied Astronomy RAS, Kutuzov embankment 10, St. Petersburg, 191187, Russia \\
  \email{dpavlov@iaaras.ru} \\ \\
  J. G. Williams \at Jet Propulsion Laboratory, California Institute of Technology,
  Pasadena, CA 91109, USA
  }
\maketitle

\begin{abstract}
The aim of this work is to combine the model of orbital and rotational
motion of the Moon developed for DE430 with up-to-date astronomical,
geodynamical, and geo- and selenophysical models. The parameters of
the orbit and physical libration are determined in this work from
lunar laser ranging (LLR) observations made at different observatories
in 1970-2013. Parameters of other models are taken from solutions that
were obtained independently from LLR.

A new implementation of the DE430 lunar model, including
the liquid core equations, was done within the EPM ephemeris.  The
postfit residuals of LLR observations make evident that the
terrestrial models and solutions recommended by the IERS Conventions
are compatible with the lunar theory. That includes: EGM2008
gravitational potential with conventional corrections and variations
from solid and ocean tides; displacement of stations due to solid and
ocean loading tides; and precession-nutation model.  Usage of these models in
the solution for LLR observations has allowed us to reduce the number
of parameters to be fit. The fixed model of tidal variations of the
geopotential has resulted in a lesser value of Moon's extra
eccentricity rate, as compared to the original DE430 model with two
fit parameters.

A mixed model of lunar gravitational potential was used, with some coefficients determined
from LLR observations, and other taken from the GL660b solution obtained from
the GRAIL spacecraft mission.

Solutions obtain accurate positions for the ranging stations and the
five retroreflectors. Station motion is derived for sites with long
data spans. Dissipation is detected at the lunar fluid core-solid
mantle boundary demonstrating that a fluid core is present. Tidal
dissipation is strong at both Earth and Moon. Consequently, the lunar
semimajor axis is expanding by 38.20 mm/yr, the tidal acceleration in
mean longitude is $-25.90 \arcsec/\mathrm{cy}^2$, and the eccentricity
is increasing by $1.48\times 10^{-11}$ each year.


\end{abstract}

\section{Introduction}
\label{intro}
Lunar laser ranging (LLR) has been the most precise way to determine the orbit and physical
libration of the Moon since 1970. Several groups across the world
issue lunar ephemerides, most notably NASA JPL \citep{williams01,folkner},
IAA RAS \citep{krasinsky02, krasinsky11, vasilyev},
and IMCCE \citep{manche07,manche10a,fienga}. Equations of motion, algorithms of reductions of observations,
and sets of determined parameters are not identical across groups, although they 
have much in common. A number of choices arise regarding parameters of gravitational potential
of Earth and Moon: one can determine them from LLR, or use preset solutions obtained
from gravimetry measurements. An additional choice is whether to determine parameters
of Earth's nutation from LLR or use a preset model with daily corrections obtained
from VLBI observations.

In this work, preset solutions are explored in the context of their compatibility with
lunar ranging observations:\begin{itemize}
\item Conventional model of geopotential and its tidal variations;
\item IAU 2000/2006 precession-nutation model with available EOP series;
\item GL660b model of lunar gravitational potential;
\item GNSS solutions for stations' drift (for selected stations).
\end{itemize}

A completely new implementation compatible with the DE430 lunar integration model was done 
on top of the implementation of the EPM ephemerides \citep{pitjeva13,pitjeva14},
along with a new implementation of reduction of LLR observations.
No original DE430 program code was used in this work.
The Moon was integrated along with the whole Solar system;
for the rest of the Solar system model, EPM's dynamical equations were used.

\section{Observations}
Observations were processed from all stations that have their LLR data publicly available.
Table \ref{tbl-observations} shows the number and timespan of observations processed from each station.

\begin{table}[h]
  \begin{tabular}{ | l | l | r | }
    \hline
    \textbf{Station} & \textbf{Timespan} & \textbf{\# of normal points} \\ \hline
    McDonald, TX, USA  & 1970--1985 & 3604 \\ \hline
    MLRS1, TX, USA     & 1983--1988 &  631 \\ \hline 
    MLRS2, TX, USA     & 1988--2013 & 3653 \\ \hline 
    Haleakala, HI, USA & 1984--1990 &  770 \\ \hline 
    CERGA, France (Ruby laser) & 1984--1986 & 1188 \\ \hline 
    CERGA, France (YAG laser)  & 1987--2005 & 8324 \\ \hline 
    CERGA, France (MeO laser)  & 2009--2013 & 654 \\ \hline
    Matera, Italy              & 2003--2013 & 83 \\ \hline
    Apache Point, NM, USA      & 2006--2012 & 1573 \\ \hline
    \textbf{total}             & \textbf{1970--2013} & \textbf{20480} \\ \hline
  \end{tabular}
  \caption{Lunar laser ranging observations available as normal points}
  \label{tbl-observations}
\end{table}

Apache Point observations were downloaded from the APOLLO website
(\url{http://physics.ucsd.edu/~tmurphy/apollo/norm_pts.html}). Observations for
the rest of the stations were downloaded from the Lunar Analysis Center of Paris Observatory
(\url{http://polac.obspm.fr/llrdatae.html}). All the downloaded files are
in so-called ``MINI'' format (one line per normal point).

Some uncertainties provided with the normal points were changed before
determination of model parameters. Uncertainties of Apache Point
observations were scaled up as recommended on the APOLLO website.
Provided normal points of Matera for the period of 4 December 2011 to
9 November 2012 have unrealistically small (few ps) uncertainties and
have been fixed to 83.4~ps (2.5~cm). For other stations, selected
groups of normal points were scaled up to match the postfit
weighted root-mean-square (wrms). Scaling was done when the postfit wrms was higher
than the rms of provided uncertainties by a 20\% margin or more. The
groups were formed following the big-picture behavior of the provided
uncertainties. The reweighting is summarized in
Table~\ref{tbl-reweight}. After refitting of the reweighted observations,
none of postfit wrms (see section~\ref{o-c}) exceeds the rms of provided
uncertainties by more than 20\%.

\begin{table}[h]
  \begin{tabular}{ | l | r | r | l | }
    \hline
    \multicolumn{1}{|c|}{\textbf{Station}} &
    \multicolumn{1}{c|}{\textbf{Timespan}} &
    \multicolumn{1}{c|}{\textbf{factor}} &
    \multicolumn{1}{c|}{\textbf{explanation}}  \\ \hline
    Haleakala & all & $\times 1.4$ & provided 4.1 cm, postfit 5.8 cm \\
    McDonald & all & $\times 1.2$ & provided 16.8 cm, postfit 20.1 cm \\
    MLRS1 & before 06.09.1985 & $\times 1.2$ & provided 34.7 cm, postfit 41.7 cm  \\
    MLRS1 & after  06.09.1985 & $\times 2.3$ & provided 2.8 cm, postfit 6.4 cm \\
    MLRS2 & before 18.06.1999 & $\times 1.6$ & provided 2.1 cm, postfit 3.5 cm \\
    CERGA & 30.10.1991--07.09.1992 & $\times 1.8$ & provided 3.1 cm, postfit 5.7 cm \\
    CERGA & 06.12.1993--01.01.1995 & $\times 1.2$ & provided 3.3 cm, postfit 4.0 cm \\
    CERGA & 01.01.1995--11.11.1998 & $\times 1.9$ & provided 1.6 cm, postfit 3.1 cm \\
   Apache & 04.04.2006--30.10.2010 & $\times 2.0$   & ``group a'' scaling by APOLLO \\
   Apache & 01.12.2010--06.04.2012 & $\times 6.0$   & ``group b'' scaling by APOLLO \\
   Apache & 07.04.2012--28.02.2012 & $\times 2.5$ & ``group c'' scaling by APOLLO \\
   Matera & 04.12.2011--09.11.2012 & to 2.5 cm & too small provided uncertainties \\ \hline
  \end{tabular}
  \caption{Reweighting of selected observations. Date notation is DD.MM.YYYY}
  \label{tbl-reweight}
\end{table}

More recent available observations were deliberately left out, in order to facilitate
possible comparison of the obtained results with already published lunar ephemerides
\citep{fienga,folkner,vasilyev}.

\section{Dynamical model}
\subsection{Planetary part}
The modeled motion of the Sun, the planets, and the Moon (as point-masses) obeys the 
Einstein-Infeld-Hoffmann relativistic equations in inertial barycentric frame and TDB timescale,
with additional perturbations from:
\begin{itemize}
\item solar oblateness;
\item 301 largest asteroids and 30 largest trans-Neptunian objects (TNO);
\item a two-dimensional asteroid annulus;
\item a one-dimensional TNO ring.
\end{itemize}
For details on planetary equations, we refer to \citep{pitjeva13,pitjeva14}
and \citep{folkner}.
The rest of this section describes the geocentric motion of the Moon and its rotation.
\subsection{Orbit of the Moon}
The following perturbations are included in the dynamical equations of the geocentric motion of the Moon:
\begin{itemize}
\item interaction between the Moon's figure and bodies considered as point masses (Earth, Sun,
  Venus, Mars and Jupiter);
\item interaction between Earth's figure and bodies considered as point masses (Moon, Sun,
  Venus, Mars and Jupiter);
\item interaction between the distorted part of the Earth (due to solid tides raised by the Moon
    and the Sun) and the Moon.
  \end{itemize}

Acceleration of a point-mass $m$ due to a body's disturbed gravitational potential
is calculated from the normalized spherical harmonic terms $\OCnm$ and $\OSnm$:
\begin{equation}\label{gravpot}
  \begin{aligned}
  \frac{\bm f_\mathrm{fig-pm}}{m} &= \mu\mathrm{Re}\left[\sum_{n=2}^{\Nmax} R^n\sum_{m=0}^n
    (\OCnm-i\OSnm)\nabla\Vnm(r,\lambda,\phi)\right]
\\
  \Vnm(r,\lambda,\phi) &=
  \Nnm \frac{\cos m\lambda + i\sin m\lambda}{r^{n+1}} P_n^m(\sin\phi)
  \\
  \Nnm &= \sqrt\frac{(n-m)!(2n+1)!(2-\delta_{0m})}{(n+m)!}
  \end{aligned}
\end{equation}
Where: $\mu$ is the body's standard gravitational parameter; $R$ is the body's radius;
$r$, $\lambda$, and $\phi$ are the distance, longitude, and latitude of the point-mass
in the body's frame;
$P_n^m$ is the associated Legendre function of degree $n$ and order $m$.
$\OCnm$ and $\OSnm$ are normalized spherical harmonic terms commonly
found in published solutions; the unnormalized terms $\Cnm = \Nnm\OCnm$
and $\Snm = \Nnm\OSnm$ were introduced in \citep{cunningham}.
The resulting acceleration should be rotated from the body's frame to
inertial frame.
We refer to \citep{krasinsky06} for the recursive equations used to calculate
$\nabla\Vnm(r,\lambda,\phi)$. $\Nmax$ is a chosen limit of the degree of expansion
of the body's gravitational potential. In this work, $\Nmax$ is 6 for both Earth and
Moon. Effects from higher degrees has proven to be unnoticeably small at the present
level of observations.

\smallskip

The Moon and the Sun raise periodical ocean and solid tides on the Earth
(there is also an additional constant distortion of $C_{20,E}$ caused by the Sun and the Moon).
Two approaches can be made to
account for the perturbations of the orbital motion of the Moon due to these
tides. Throughout this paper, we will reference to them as the ``IERS tidal model''
and the ``DE tidal model''.

\subsubsection{IERS tidal model: variations of spherical harmonic coefficients}
\label{de-tidal-model}
It is recommended by the IERS Conventions \citep{iers2010} that the
changes induced by the solid and ocean tides are modeled as variations
in the coefficients $\bar{C}_\mathrm{nm}$ and $\bar{S}_\mathrm{nm}$. Only
corrections up to order and degree 2 are taken. Solid tide
corrections for ``conventional tide free'' EGM2008 are computed in two
steps. At first the frequency-independent part is computed: 
\begin{equation}
\Delta\bar{C}_\mathrm{nm,E}-i\Delta{\bar{S}_\mathrm{nm,E}}= \frac{k_\mathrm{nm}}{2n+1}
\sum_{j=M,S}\frac{\mu_j}{\mu_E}\left(\frac{R_E}{r_j}\right)^{n+1}
\bar{P}_\mathrm{nm} (\sin\Phi_j)e^{-im\lambda_j}
\label{eq:norm_geo_coeffs} \end{equation}
where $\mu_E$, $\mu_M$, and $\mu_S$ are the standard gravitational parameters of the Earth, Moon,
and Sun respectively; $\bar{P}_\mathrm{nm}=P_n^m N_\mathrm{nm}$ is the normalized associated Legendre
polynomial, and the Love numbers $k_\mathrm{nm}$ correspond to those $\mathrm{(nm)}$
coefficients being corrected.  Since elastic properties of the Earth
are frequency dependent, on the second step one should compute
additional corrections from the respective bands to the coefficients
using frequency dependent Love numbers different from respective
nominal values. The correction for $\bar{C}_{20,E}$ from the long period
components is:
\begin{equation}
\Delta\bar{C}_{20}^\mathrm{(fd)} = {\rm Re} \sum_f(A_0\delta k_f H_f) e^{i{\theta_f}}
\label{eq:c20_stide} \end{equation}
and the corrections to $\bar{C}_\mathrm{2m}$ and $\bar{S}_\mathrm{2m}$
from diurnals ($m=1$) and semidiurnals ($m=2$) are given by
\begin{equation}
\begin{cases}
  \Delta\bar{C}_{21}^\mathrm{(fd)}-i\Delta\bar{S}_{21}^\mathrm{(fd)} =
  -i \sum\limits_f(A_1\delta k_f H_f) \,e^{i{\theta_f}}\\
\Delta\bar{C}_{22}^\mathrm{(fd)}-i\Delta\bar{S}_{22}^\mathrm{(fd)} = \sum\limits_f(A_2\delta k_f H_f) \,e^{i\theta_f}
\end{cases},
\label{eq:cs2m_stide} \end{equation}
where
\begin{equation}
\begin{aligned}
  A_0 &= \frac{1}{R_E\sqrt{4\pi}}, \\
  A_m &=  \frac{(-1)^m}{R_E\sqrt{8\pi}}, \qquad (m=1,2).
  \end{aligned}
\end{equation}

$\theta_f(t) = \bar{n} \cdot \bar{\beta}(t)$ is the argument of respective tide constituent $f$.
$\delta k_f = \delta k^R_f+i\delta k^I_f$ is the difference for a Love
number from its nominal value on frequency $f$. $H_f$ is the amplitude
of the term on $f$, $\bar{\beta}=(\tau, s, h, p, N', p_s)$ is a six-vector of Doodson's
fundamental arguments, $\bar{n}$ is a six-vector of multipliers of the fundamental arguments,
and 
$\mathrm{(fd)}$ denotes ``frequency dependent''.  The detailed information about
these terms and their computation is given in \citep[ Chapters 5 and 6]{iers2010}.

Corrections to Stokes coefficients to account for effects of the ocean tides are expressed
as
\begin{equation}
\Delta\bar{C}_\mathrm{nm}^\mathrm{(ocean)}-i\Delta\bar{S}_\mathrm{nm}^\mathrm{(ocean)} =
\sum_{f} \sum_+^- ({\cal C}_\mathrm{f,nm}^\pm \mp i {\cal S}_\mathrm{f,nm}^\pm) e^{\pm i{\theta_f}},
\label{eq:norm-coeff} \end{equation}
where $\theta_f = m(\theta_g + \pi) -  \bar{N} \cdot \bar{F}$,
$\bar{F}=(l,l',F,D,\Omega)$ is a five-vector of Delaunay variables  of nutation theory,
$\bar{N}$ is a five-vector of multipliers of the Delaunay variables for the nutation of
frequency $(-f+\mathrm{d}\theta_g/\mathrm{d}t)$, $\theta_g$ is GMST (in angle units). 
${\cal C}_\mathrm{f,nm}^\pm$ and ${\cal S}_\mathrm{f,nm}^\pm$ are the
harmonic coefficients of the main waves of the ocean tides model
FES2004 recommended for use by the IERS Conventions 2010. Their values can
be taken from
\url{http://tai.bipm.org/iers/convupdt/convupdt_c6.html}. Detailed
information about the effect of ocean tides on the geopotential is
given in \citep[ Section 6.3]{iers2010}.

\subsubsection{DE tidal model: direct acceleration with five time delays}
\label{iers-tidal-model}
The full description of the  model used in the DE430 ephemeris can be found
in \citep[ Section III.C]{folkner}. The acceleration of the Moon is evaluated
separately for the tides raised by the Sun and the Moon itself, on three frequencies:
zonal (i.e. due to variation of{ $C_{20,E}$), diurnal ($C_{21,E}$ and $S_{21,E}$),
  and semi-diurnal ($C_{22,E}$ and $S_{22,E}$).
Each of the three frequencies has its fixed Love number $k_{2m}$.
Tidal dissipation causes the response of the earth to be delayed. Consequently,
the perturbing acceleration from a tide-raising body at order $m$ at
time $t$ is derived from Eq. (\ref{gravpot}) using tidal response
$\Delta \bar C_\mathrm{2m,E}$, $\Delta \bar S_\mathrm{2m,E}$ created by the body at time $t-\taumO$
at the Earth rotated back to time $t-\taumR$.
Pragmatically, the terrestrial phase shifts depend on tidal period and the two extra
delays $\tau_\mathrm{1O}$ and $\tau_\mathrm{2O}$ allow the diurnal and semidiurnal tidal
phases to vary linearly with frequency \citep{williams16}.

We denote $\bm r(t)$ the geocentric position of the Moon.
The tidal distortion for each order is computed by replacing
the geocentric position of the tide-raising body $\bm r_\mathrm{body}(t)$ with
$\bm{r}_m^*=R_z(\dot\theta_E \taumR)\bm r_\mathrm{body}(t-\taumO)$, where
$\dot\theta_E$ is the Earth's sidereal rotation rate.  We break down the vectors to
``equatorial'' and ``polar'' components with respect to the Earth's equator:
$\bm r = \bm\rho + \bm z$, $\bm{r}_m^* =  \bm\rho_m^* + \bm z_m^*$.
Parameters with an asterisk are used for calculating the tide.
The equation for the perturbing acceleration
of the Moon is:
\begin{equation}\label{detides}
  \begin{aligned}
    \frac{\Delta \bm f}{m} = \frac{3\mu_j}{2}\left(\frac{R_E}{r}\right)^5 &
    \left[\frac{k_{20}}{{r_0^*}^5} \left(\left(2{z_0^*}^2\bm z + {\rho_0^*}^2\bm\rho\right)
      - 5\frac{(zz_0^*)^2+\frac{1}{2}(\rho\rho_0^*)^2}{r^2}\bm r + {r_0^*}^2\bm r \right)\right.\\
      & + \frac{k_{21}}{{r_1^*}^5}\left( 2\left( (\bm\rho\cdot\bm\rho_1^*)\bm z_1^* +
      zz_1^*\bm\rho_1^*\right) -
      \frac{10zz_1^*(\bm\rho\cdot\bm\rho_1^*)\bm r}{r^2} \right)\\
      & \left. + \frac{k_{22}}{{r_2^*}^5}\left( 2(\bm\rho\cdot\bm\rho_2^*)\bm\rho_2^* - {\rho_2^*}^2\bm\rho
        - 5\frac{(\bm \rho \cdot \bm \rho_2^*)^2 - \frac{1}{2}(\rho\rho_2^*)^2 }{r^2} \bm r
        \right) \right],
    \end{aligned}
  \end{equation}
where $\mu_j$ is the gravitational parameter of the tide-raising body, and
$R_E$ is Earth's equatorial radius.
The acceleration is given in the inertial frame for one tide-raising body; to get the total
perturbing geocentric acceleration of the Moon, one has to add up the results of Eq. (\ref{detides})
with the Moon and the Sun as the tide-raising bodies, and then multiply by $(1+\mu_M/\mu_E)$.

Zonal tides do not depend on the rotation of the Earth, so $\tau_{0R}=0$. Other rotational
delays, $\tau_\mathrm{1R}$ and $\tau_\mathrm{2R}$, are determined from observations. Love numbers and
orbit delays are fixed to match the most influential solid Earth tides and ocean tides
from known models. Values used in this work are $k_{20}=0.335$, $k_{21}=0.320$,
$k_{22}=0.282$, $\tau_\mathrm{0O}=0.0780$~d, $\tau_\mathrm{1O}=-0.044$~d, and $\tau_\mathrm{2O}=-0.113$~d.
Those values are modified from the ones used in DE430, but note $k_{22}\tau_\mathrm{2O}=-0.031866$
is about the same as for DE430 ($0.320 \times -0.1 = -0.032$).

The negative values of $\tau_\mathrm{mO}$ reflect the increase of ocean phase shift with period
rather than a response to the future position of the Moon. The negative $\tau_\mathrm{1O}$ reflects
the increase in the diurnal phase lag between the O1 and Q1 tides, while the negative
$\tau_\mathrm{2O}$ reflects the increase in the semidiurnal phase lag between the M2 and N2 tides.
For details, we refer to \citep[ Section 4]{williams16}.

\subsection{Lunar frame}
The lunar frame is aligned with the principal axes of the undistorted lunar mantle.
The orientation of the lunar frame w.r.t. the inertial frame is determined by three Euler angles:
$\phi$, $\theta$, and $\psi$ that evolve over time.
The transformation from the lunar frame to the inertial frame is given by the matrix:
\begin{equation}
  R_{L2C}(t)=R_z(\phi(t))R_x(\theta(t))R_z(\psi(t)).
\end{equation}
$R_x$ and $R_z$ are matrices of right-hand rotations around axes $x$ and $z$, respectively.
The argument $t$ will be omitted when appropriate.

Instantaneous rates of the Euler angles at time $t$ are denoted $\dot\phi(t)$, $\dot\theta(t)$, and
$\dot\psi(t)$. Let $\bm{\omega}(t)$ be the angular velocity of the mantle,
referred to the lunar frame:
\begin{eqnarray}
  \omega_x &=& \dot\phi \sin\theta \sin\psi + \dot\theta \cos\psi \nonumber \\
  \omega_y &=& \dot\phi \sin\theta \cos\psi - \dot\theta \sin\psi \\
  \omega_z &=& \dot\phi \cos\theta + \dot\psi \nonumber
\end{eqnarray}
The behavior of the lunar mantle depends on $\bm{\dot\omega}(t)$ and obeys the following second
derivatives of the Euler angles \citep{standish}:
\begin{eqnarray}
  \ddot\phi &=& \frac{\dot\omega_x\sin\psi + \dot\omega_y\cos\psi + \dot\theta
    (\dot\psi - \dot\phi\cos\theta)}
            {\sin\theta} \nonumber \\
  \ddot\theta &=& \dot\omega_x\cos\psi - \dot\omega_y\sin\psi - \dot\phi\dot\psi\sin\theta \\
  \ddot\psi &=& \dot\omega_z - \ddot\phi\cos\theta + \dot\phi\dot\theta\sin\theta \nonumber
\end{eqnarray}
$\bm{\dot\omega}(t)$, in turn, depends on the torque $\bm N(t)$. Using the Euler's equation for the
angular momentum in a rotating reference frame (${\bm N} = {\bm{\dot{L}}} + \bm\omega\times{\bm L}$, where
$\bm L = I\bm\omega$ is the angular momentum in the rotating frame), we can write $\bm{\dot\omega}(t)$
in the following form:
\begin{equation}
  \bm{\dot\omega} = \left(\frac{I}{m}\right)^{-1}\left[\frac{\bm{N}}{m} - \frac{\dot{I}}{m}\bm\omega - \bm\omega\times\left(\frac{I}{m}\bm\omega\right)\right]
\end{equation}
where $m$ is the mass of the Moon, and $I(t)$ is the inertia tensor of the lunar mantle.
The torque $\bm N$ (also referred to the lunar frame) is calculated as:
\begin{equation}
  \frac{\bm N}{m} = \sum\limits_{A\neq M}\frac{{\bm N}_\mathrm{figM-pmA}}{m} +
                  \frac{{\bm N}_\mathrm{figM-figE}}{m} + \frac{\Ncmb}{m},
\end{equation}
where ${\bm N}_\mathrm{figM-pmA}(t)$ is a torque from point-mass $A$ to the Moon's figure:
\begin{equation}
\frac{{\bm N}_\mathrm{figM-pmA}}{m} = (\bm{r}_M-\bm{r}_A)\times \frac{{\bm f}_\mathrm{figM-pmA}}{m},
  \end{equation}
where ${\bm f}_\mathrm{figM-pmA}(t)$ is the force acting on the point-mass in the Moon's gravitational
field (see Eq. \ref{gravpot}). The following point-masses are taken into account: Earth, Sun, Venus, Mars, Jupiter.

${\bm N}_\mathrm{figM-figE}(t)$ is a torque from the Earth's oblateness to the Moon's figure:

\begin{equation}\begin{aligned}
  \frac{{\bm N}_\mathrm{figM-figE}}{m} = \frac{15\mu_E R_E^2 J_\mathrm{2E}}{2r_\mathrm{EM}^5}& \left[
    \left(1-7(\rhatEM\cdot\phat)^2\right)
    \left(\rhatEM \times \frac{I}{m} \rhatEM\right)\right. \\ &  
     + 2 (\rhatEM \cdot\phat)
    \left(\rhatEM \times \frac{I}{m} \phat + \phat \times \frac{I}{m}\rhatEM \right) \\ &
    \left.  - \frac{2}{5}\left(\phat \times \frac{I}{m} \phat\right)\right],
\end{aligned}\end{equation}
where $J_\mathrm{2E}$ is Earth's oblateness factor, $r_\mathrm{EM}(t)$ is Earth-Moon distance,
$\rhatEM(t)$ is the normalized direction vector from the Moon to the Earth, and $\phat(t)$ is
the direction of Earth's pole. All vectors in the equation are referred to the lunar frame.

$\Ncmb$ will be explained in section \ref{cmb-subsection}.

\subsection{Lunar inertia tensor}\label{sec-inertia}
The inertia tensor of the lunar mantle is subject to delayed
tidal distortion from Earth and delayed spin distortion.
We refer to \citep{williams01} and \citep{folkner} for full descriptions,
while reproducing the equation here in its condensed form:
\begin{equation}
  \label{inertia-tensor}
  \begin{aligned}
    \frac{I}{m} & = \frac{2R_M^2\tilde J_2}{2\beta-\gamma+\beta\gamma}\left[
      \begin{array}{ccc}1-\beta\gamma & 0 & 0 \\
        0 & 1+\gamma & 0 \\
        0 & 0 & 1+\beta \end{array}\right] - \frac{I_c}{m} \\
    & - k_2\frac{\mu_E}{\mu_M}\left(\frac{R_M}{r}\right)^5
    \left[\begin{array}{ccc}
        x^2-\frac{1}{3}r^2 & xy & xz \\
        xy & y^2-\frac{1}{3}r^2 & yz \\
        xz & yz & z^2-\frac{1}{3}r^2
      \end{array}\right] \\
    & + k_2\frac{R_M^5}{3\mu_M}\left[\begin{array}{ccc}
        \omega_x^2-\frac{1}{3}(\omega^2-n^2) & \omega_x\omega_y & \omega_x\omega_z \\
        \omega_x\omega_y & \omega_y^2-\frac{1}{3}(\omega^2-n^2) & \omega_y\omega_z \\
        \omega_x\omega_z & \omega_y\omega_z & \omega_z^2 - \frac{1}{3}(\omega^2+2n^2) \\
        \end{array}\right],
    \end{aligned}
  \end{equation}
where
$R_M$ is the equatorial radius of the Moon;
$\tilde J_2$ is the oblateness factor of the undistorted Moon;
$k_2$ is the degree-2 Love number of the Moon; 
$\bm r = (x, y, z)^T$ is the position of the Moon relative to Earth; $n$ is the lunar mean motion.
$I_c(t)$ is the inertia tensor of the liquid core that is explained in section \ref{cmb-subsection}.
Tidal and spin distortions are evaluated with a delayed argument: calculation of $I(t)$
involves not $\bm r(t)$ and $\bm\omega(t)$, but $\bm r(t-\tau)$ and $\bm\omega(t-\tau)$.

The distortion of the lunar mantle affects its gravitational potential. The following
equations describe how unnormalized spherical harmonic coefficients vary over time:
\begin{equation}\label{dynamical-potential}
  \begin{aligned}
    C_{20} & = \frac{1}{R_M^2}\left[\frac{1}{2}\left(\frac{I_{11}^{*}}{m}+
                  \frac{I_{22}^{*}}{m}\right)-\frac{I_{33}^{*}}{m}\right] \\
    C_{22} & = \frac{1}{4R_M^2}\left[\frac{I_{22}^{*}}{m}-\frac{I_{11}^{*}}{m}\right] \\
    C_{21} & = C_{21}^{(0)} - \frac{1}{R_M^2}\frac{I_{13}^{*}}{m} \\
    S_{21} & = S_{21}^{(0)} - \frac{1}{R_M^2}\frac{I_{32}^{*}}{m} \\
    S_{22} & = S_{22}^{(0)} - \frac{1}{2R_M^2}\frac{I_{21}^{*}}{m}
\end{aligned} \end{equation}
Here the matrix $I^{*}$ is the combined inertia tensor: $I^{*}=I+I_c$.
The DE430 lunar equations are built on assumption that the mantle frame is aligned with
the principal axes of the whole Moon, so the constant mean values
$C_{21}^{(0)}$, $S_{21}^{(0)}$, and $S_{22}^{(0)}$ should be zero. $I^{*}$ then
is the total inertia tensor of the Moon. However, in this work a preliminary attempt
has been made to include a nonzero $S_{21}^{(0)}$, see section~\ref{borderline-subsec}.

\subsection{Lunar fluid core}\label{cmb-subsection}
The core is assumed to be rotating like a solid and constrained by the shape of the core-mantle
boundary (CMB) at the interior of the mantle, with moment of inertia constant
in the mantle frame \citep{folkner}:
\begin{equation}
  \frac{I_c}{m} = \alpha_c \frac{C_T}{m}\left[\begin{array}{ccc}
      1-f_c & 0 & 0 \\
      0 & 1-f_c & 0 \\
      0 & 0 & 1
      \end{array}
      \right], \quad \frac{C_T}{m} = \frac{2(1+\beta)}{2\beta-\gamma+\beta\gamma}R_M^2\tilde J_2,
  \end{equation}
where $\alpha_c$ is a dimensionless coefficient for the ratio of core to total polar moments of inertia and $f_c$ is the oblateness of the fluid core.
$C_T$ is the undistorted polar moment of inertia.

The orientation of the core is not important
for the equations of the mantle---only its angular velocity $\bm\omega_c(t)$ is. The evolution of
$\bm\omega_c$ is described by its time derivative, given in the mantle frame:
\begin{equation}
  \bm{\dot\omega_c} = \left(\frac{I_c}{m}\right)^{-1} \left[-\bm\omega\times \frac{I_c}{m}\bm\omega_c -
    \frac{\Ncmb}{m} \right]
\end{equation}

$\Ncmb(t)$ is the torque on the mantle due to the interaction with the fluid core. It is expressed
in the mantle frame as:
\begin{equation}
  \frac{\Ncmb}{m} = \frac{C_T}{m} \left[\frac{k_v}{C_T}(\bm\omega_c - \bm\omega) +
  \alpha_c f_c (\bm{\hat z}\cdot\bm\omega_c)(\bm{\hat z}\times\bm\omega_c)\right],
  \end{equation}
where $\frac{k_v}{C_T}$ is a friction parameter (measured in day$^{-1}$), and $\bm{\hat z}=(0, 0, 1)^T$.

\section{Reductions of observations}
The calculation of modeled light travel time requires solving a system of equations:
\begin{equation}\label{traveltime}
\begin{cases}
t_2 - t_1 = \frac{|\LBCRS(t_2) - \SBCRS(t_1)|}{c} +
              \Dgrav(t_1, t_2) + \Datm(t_1, t_2) \\
  t_3 - t_2 = \frac{|\SBCRS(t_3) - \LBCRS(t_2)|}{c} +
              \Dgrav(t_3, t_2) + \Datm(t_3, t_2)
\end{cases}
\end{equation}

$t_1$, $t_2$, and $t_3$ are the times of emission, reflection, and reception of the signal
in the TDB timescale. Usually, a normal point contains $t_1$ in UTC, which requires converting
it to TDB and then solving (\ref{traveltime}) w.r.t. $t_2$ and $t_3$.
$\SBCRS(t_i)$ and $\LBCRS(t_i)$ are the positions of the station and the lunar reflector
at time $t_i$ in the inertial frame. $\Dgrav$ is the relativistic gravitational delay
of signal propagation, while $\Datm$ is the tropospheric delay.

Calculation of the position of the station in the inertial frame should
include a relativistic transformation from geocentric to barycentric coordinate systems
\citep[ eq. 11.19]{iers2010}:
\begin{equation}
  \SBCRS = \bm r_E + \SGCRS\left(1-\frac{U_E}{c^2} - L_C\right) -
     \frac{1}{2}\left(\frac{\bm{\dot r}_E\cdot \SGCRS}{c^2}\right)\bm{\dot{r}}_E
\end{equation}
where $U_E(t)$ is the gravitational potential at the geocenter, excluding the Earth's mass,
$\bm{r}_E(t)$ and $\bm{\dot{r}}_E(t)$ are the barycentric position and velocity of the Earth,
and $L_C=1.48082686741\times 10^{-8}$.

Calculation of inertial geocentric position of the station should 
account for Earth's rotation, pole tides, and solid body and ocean loading tides:
\begin{equation}
  \SGCRS = R_\textrm{T2C}\left(\STRS + \bm\Delta_\mathrm{pole} + \bm\Delta_\mathrm{solid} +
                                 \bm\Delta_\mathrm{ocean}\right).
\end{equation}
where $\STRS(t)$ is the position of the station in the terrestrial reference frame,
adjusted for the drift.

Note: the Haleakala station had its receiving telescope ($\STRS(t_3))$
located at some distance from the laser ($\STRS(t_1)$), so that there
are two different $\STRS$ in the equations. The displacement between
the telescope and the laser is not determined from the observations;
the position of the laser is determined, while the position of the telescope
is calculated from its known displacement that can be found in \citep{newhall}.

The transformation $R_\textrm{T2C}$ from the TRS to GCRS frame is done strictly in accordance
with IAU 2000/2006 resolutions. The algorithms for the transformation can be found in
\citep[ Chapter 6]{iers2010}. Unmodeled celestial pole offsets $\mathrm{d}X(t)$ and
$\mathrm{d}Y(t)$ and terrestrial pole coordinates $(x_\mathrm{pole}(t), y_\mathrm{pole}(t))$
were taken from the published IERS C04 solution \citep{bizouard09,bizouard11} 
which is combined from SLR, GPS, and VLBI data, including the observations from the
QUASAR network \citep{finkelstein}.
However, the C04 solution gave poor results for observations made before 1982
(the only station from which we have data for that time is McDonald). The decision
has been made to use the JPL KEOF series (\url{http://keof.jpl.nasa.gov/}) for
early observations. Figure~\ref{mcdonald-keof} shows the postfit residuals
of the McDonald observations obtained using C04 (one-way wrms 30.8 cm) and KEOF
(one-way wrms 20.1 cm).
\begin{figure}[h]
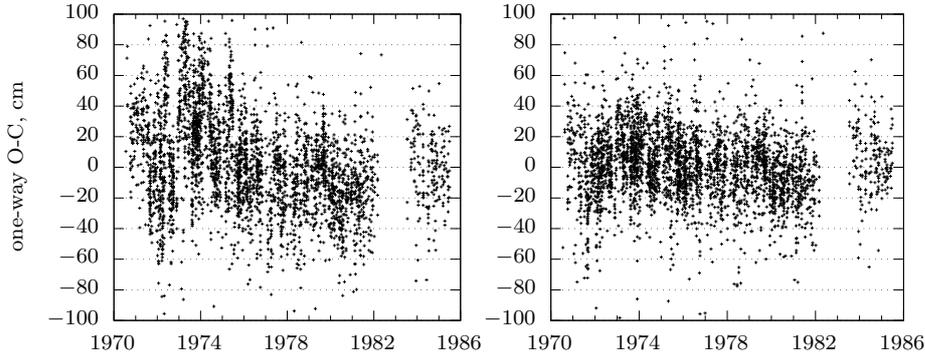

\centering
\caption{Postfit O-C of the McDonald observations using C04 (left) and KEOF (right)}
\begin{minipage}{0.49\linewidth}
\input residuals-mcdonald-nokeof.tex
\end{minipage}
\begin{minipage}{0.49\linewidth}
\input residuals-mcdonald-keof.tex
\end{minipage}
\label{mcdonald-keof}
\end{figure}

The reason for KEOF giving better results can be
that the variation of latitude (VOL) and UT0 determined from LLR
observations were part of the KEOF solution \citep{ratcliff};
besides, the C04 series has $\mathrm{d}X=\mathrm{d}Y=0$ before 1984.
Since 1984, C04 and KEOF give equally good results for all stations
(there is almost no LLR data between 1982 and 1984). Figure \ref{keof-c04-ut1},
showing the difference  between the KEOF and C04 series of UT1, confirms
that the two solutions came close enough to each other starting around 1984.
\begin{figure}[h]
  \caption{$\mathrm{UT1}_\mathrm{KEOF}-\mathrm{UT1}_\mathrm{C04}$, seconds}
  \input keof-c04-ut1.tex
  \label{keof-c04-ut1}
  \end{figure}

For $\bm\Delta_\mathrm{pole}(t)$, $\bm\Delta_\mathrm{solid}(t)$, and $\bm\Delta_\mathrm{ocean}(t)$,
we refer to the respective sections of \citep{iers2010}:
7.1.1 (solid Earth tide), 7.1.2 (ocean loading tide),
and 7.1.4 (pole tide). Atmospheric pressure
loading and ocean pole tide loading are not handled in this work.

\smallskip

The position of the lunar reflector has to be transformed from
the lunar frame to the inertial frame, similar to the position of the station:
\begin{equation}
  \begin{aligned}
  \LBCRS &= \bm r_M + \LLCRS\left(1-\frac{U_M}{c^2}\right) -
  \frac{1}{2}\left(\frac{\bm{\dot r}_M\cdot \LLCRS}{c^2}\right)\bm{\dot{r}}_M \\
  \LLCRS &= R_\textrm{L2C}\,\LPA + \bm\Delta_\mathrm{solidmoon}^{(E)} + \bm\Delta_\mathrm{solidmoon}^{(S)}
    \end{aligned}
\end{equation}
where $U_M(t)$ is the gravitational potential at the Moon's center, excluding the Moon's mass,
$\bm{r}_M(t)$ and $\bm{\dot{r}}_M(t)$ are the barycentric position and velocity of the Moon,
$\LPA$ is the position of the reflector in the lunar frame (principal axes), and
$\bm\Delta_\mathrm{solidmoon}^{(E)}(t)$ and $\bm\Delta_\mathrm{solidmoon}^{(S)}(t)$ are
displacements due to solid Moon tide raised by Earth and Sun, respectively.
A simple model of solid Moon tides was used in this work, while more detailed
models have been recently developed; see \citep{williams15}.

The equation of the tide involves the degree-2 Love number $h_2$ and the
degree-2 Shida number $l_2$ \citep[ eq. 7.5]{iers2010}:
\begin{equation}
  \begin{aligned}
  \bm\Delta_\mathrm{solidmoon} = \frac{\mu_A R_M^4}{\mu_M r_\mathrm{MA}^3}
  \left[\vphantom{\frac{h_2}{2}}\right. &
     \frac{h_2}{2}\left(3\rMAcdotl^2 - 1\right)\bm{\hat{r}}_\mathrm{MA} +  \\
     & \left. 3 l_2 \rMAcdotl \left(\bm{\hat{r}}_\mathrm{MA} - \rMAcdotl \bm{\hat l}\right)
     \vphantom{\frac{h_2}{2}}\right],
  \end{aligned}
  \end{equation}
where $\bm{\hat{l}} = R_\textrm{L2C}\,{\bm{\hat{l}}_\mathrm{PA}}$ is the unit vector of the
reflector rotated to the inertial frame, $\bm r_\mathrm{MA} = \bm r_\mathrm{A} - \bm r_\mathrm{M}$
is the position of the tide-raising body relative to the Moon, and $\bm{\hat r}_\mathrm{MA}$
is the respective unit vector.

\smallskip

Calculation of $\Datm$ is done using a combination of two empirical models: zenith delay
(Mendes and Pavlis 2004) and mapping function \citep{mendes02}.
For the calculation of $\Dgrav$, a theoretical result is used that can be found
for instance in \citep{kopeikin}. Delays from the following point-masses are added up:
Sun, Earth, Moon, Jupiter, Saturn.

\smallskip

The observed ranges are given in UTC timescale, so the resulting ``computed'' observation should
be transformed from the TDB timescale to TT and then to UTC.
From February 1968 till the end of 1971, UTC ran faster than TT by the factor of ($1+3\times10^{-8}$).
Since 1972, UTC and TT have the same rate, while UTC has jumps. Given that the earliest LLR observations
were made in 1969, and that no LLR normal point has $t_1$ and $t_3$ on different sides
of a UTC jump, we can assume that $(t_3-t_1)^\textrm{(UTC)}=(t_3-t_1)^\textrm{(TT)}/(1+\zeta)$,
where $\zeta=3\times10^{-8}$ before 1972 and zero since 1972.
The complete transformation from TDB to UTC will be:
\begin{equation}\label{computed}
  \begin{aligned}
    C =  \left[t_3 - t_1\vphantom{x^x}\right. + &\ \textrm{TTminusTDB}(t_3, \SGCRS(t_3))\ -\\
      & \left.\vphantom{x^x}\textrm{TTminusTDB}(t_1, \SGCRS(t_1))\right]/\,(1+\zeta) +
    \frac{b}{c}
  \end{aligned}
  \end{equation}
For calculating $(\textrm{TT}-\textrm{TDB})$ at time $t$ and
point $\SGCRS(t)$, a theoretical equation is used, which can
be found for instance in \citep[ eq. 5]{folkner}. The geocentric terms of the equation are integrated
along with the Solar system equations and stored in ephemeris; just one topocentric term
is taken into account in Eq. (\ref{computed}): $\left(\bm{\dot r}_E(t)\cdot\SGCRS(t)\right)/c^2$.

The bias $b$ is a determined parameter and is specific to a station and to a certain
period of time. The list of biases applied in this work can be found in section
\ref{biases-sec}.

\section{Determined and fixed parameters}\label{parameters-sec}
Some of the parameters used in this model, like the positions of the lunar reflectors,
are to be determined from LLR exclusively. Other parameters, like the ones
of the orientation of the Earth, are assumed to be determined from VLBI and GNSS
observations with better accuracy than they could have possibly been determined
from LLR.
\subsection{Borderline parameters}\label{borderline-subsec}
There are choices regarding parameters that can be 
determined either from LLR or alternative techniques. Such parameters are:
\begin{itemize}
\item Spherical harmonics of the lunar gravitational potential can be determined from
  LLR or from the observations made during the GRAIL spacecraft mission \citep{konopliv}
\item the mass of the Moon can be determined from LLR (given the mass of the Earth
  from some other solution), or from GRAIL.
\item tidal parameters of the Moon ($h_2$, $k_2$, $l_2$) can be determined from LLR or
  taken from GRAIL \citep{williams14} or other solutions.
\item parameters of tidal variations of the Earth's gravitational potential can be
  determined from LLR or gravimetry and altimetry measurements (the latter is the basis of
  the model recommended in the IERS Conventions).
\item drift of the stations can be determined from LLR or GNSS observations.
  \end{itemize}

Lunar $\tilde J_2$, $k_2$ and $l_2$ in this work were fixed to the values determined from GRAIL,
while $h_2$ was determined from LLR. $\mu_E+\mu_M$ was determined from LLR too, while
$\frac{\mu_E}{\mu_M}$ was fixed to the value determined from spacecraft observations.

GRAIL's undistorted value of $\bar C_{22}$ was left out: following Eq. (\ref{dynamical-potential}),
$C_{22}(t)$ is calculated dynamically with $\beta$ and $\gamma$ as determined parameters.
Undistorted (mean) values $C_{21}^{(0)}$, $S_{21}^{(0)}$, and $S_{22}^{(0)}$ are fixed to zero in DE430,
so that the mantle frame is aligned with the principal axes. The nonzero values of
$\bar C_{21}$, $\bar S_{21}$, and $\bar S_{22}$ in the GRAIL's solution are a sign of misalignment
caused by inner structure of the Moon. Currently, there is no model explaining this
misalignment, and since the present theory has Eqs. (\ref{inertia-tensor})
and~(\ref{dynamical-potential}) in the PA frame, it makes little sense to just set
$C_{21}^{(0)}$, $S_{21}^{(0)}$, and $S_{22}^{(0)}$ based on the use GRAIL's values 
$\bar C_{21}$, $\bar S_{21}$, and $\bar S_{22}$. However, a separate solution was
obtained in this work, to test how the dynamical system behaves with 
$S_{21}^{(0)}$ taken from GL660b.

The physical processes in the lunar core, mantle, and the core-mantle
boundary seem to have yet-unmodeled effects that can be presented in
the form of empirical correction of lunar gravitational potential coefficients.
By trial and error, it has been found that fitting $\bar C_{32}$, $\bar S_{32}$, and $\bar C_{33}$
to observations gives better results.

The DE430 and IERS 2010 tidal models (see subsections \ref{de-tidal-model} and
\ref{iers-tidal-model}) were both implemented in this work, and a solution
was obtained with each.

\smallskip

The drift of stations was modeled as linear motion in a cylindrical coordinate system
$(\lambda, r\cos\phi, r\sin\phi)$. The choice of the coordinate system was historical;
while it is not strictly consistent with the IERS Conventions, where
tectonic plate motions are modeled with linear function in cartesian coodinates,
the nonlinearities on relatively short timespans, used in this work, are very small
and can be ignored.

For stations that have been doing LLR for decades---McDonald/MRS1/MLRS2 and CERGA---
the velocities were fit to the observations. Velocities of Haleakala and Matera
can not be determined from LLR with good confidence, as they have relatively short
timespans of LLR observations; but they are equipped with GPS receivers, which
allowed taking their velocities from a global GNSS solution for the terrestrial frame.

The International GNSS Service (IGS) provides weekly combined coordinate
solutions for IGS stations network \citep{ferland}.
Every solution is the result of a combination of independent estimates
of solutions provided by different IGS Analysis Centers. Coordinates
are aligned to IGS realizations of ITRF. Details and links to data are
available at IGS website
\url{http://igscb.jpl.nasa.gov/components/prods.html}. To get
velocities of stations we have fit all of the coordinate time-series
for considered stations to a linear model of movement.

The Apache Point station is not a part of any ITRF solution (it does not
possess a GPS receiver). The closest station to Apache Point that is present
in the ITRF2014 solution is White Sands (WSMN, 65 km away); but it
is located on the desert floor, while Apache Point is in the mountains.
The desision has been made to take the velocity of a GPS station P027
(2.5 km away from Apache Point) from a PBO solution given in the IGS08
frame (\url{ftp://data-out.unavco.org/pub/products/velocity/pbo.final_igs08.vel}).
In future work, when more Apache Point observations are processed,
detecting its velocity from LLR observations can be considered.

\subsection{Special parameters for unmodeled effects}
\subsubsection{Longitude libration}\label{sec-long-libr}
The DE430 lunar theory includes three additional periodic terms for longitude libration
to account for small effects related to frequency dependent tidal dissipation
\citep{williams13}.
\begin{equation}\label{libr-long}
  \Delta\Lambda = A_1 \cos l' + A_2 \cos(2l-2D) + A_3\cos(2F-2l).
\end{equation}
The equation involves Delaunay arguments: lunar mean anomaly $l$,
solar mean anomaly $l'$, argument of latitude $F$, and elongation of Moon from Sun $D$.
$A_1$, $A_2$, and $A_3$ are the special parameters to be determined from observations.

The lunar mantle is supposed to have an unmodeled libration in longitude
by the periodic $\Lambda(t)$ in the MER (mean Earth -- mean rotation) frame.
This is equal to the following rotation in our chosen PA (principal axes) frame:
\begin{equation}
  R_\mathrm{libr}(\Lambda) = R_x(-\delta_x) R_y(-\delta_y) R_z(\Lambda) R_y(\delta_y) R_x(\delta_x),
\end{equation}
where constant angles $\delta_x$ and $\delta_y$ are derived from an ephemeris
to match the transformation from the MER frame to the PA frame. In DE430
\citep{folkner}, $\delta_x = 0.285\arcsec$ and $\delta_y = 78.580\arcsec$.
In this work, a simplification has been made: $R_\mathrm{libr}(\Lambda) \approx R_z(\Lambda)$,
since the change of axis (PA Z instead of MER Z) brings just sub-millimeter
differences of calculated ranges. Thus, the total lunar rotation matrix becomes
$R_\mathrm{L2C} = R_z(\phi)R_x(\theta)R_z(\psi+\Lambda)$.

\subsubsection{Extra eccentricity rate}
Tidal dissipation effects in Earth and Moon cause a secular growth of eccentricity
of the orbit of the Moon.
The eccentricity rate derived from DE430 \citep{williams13} is $1.36\times 10^{-11}$~/yr.
An extra eccentricity rate is determined to detect unmodeled (tidal or other) effects
in the orbit of the Moon. From \citep{chapront}, we know the effect on the Earth-Moon
distance $A(t)$ with the Delaunay arguments:
\begin{equation}
  A = 385000.5 - 20905.4 \cos l - 3699.1 \cos(2D-l) - \dots \ \mathrm{km.}
  \end{equation}
Since the terms with the $l$ argument have a hidden $e$ (eccentricity) in the coefficient, we can derive
\begin{equation}
\mathrm{d}A/\mathrm{d}e \approx - \frac{20905.4}{e}  \cos l - \frac{3699.1}{e} \cos(2D-l) \approx
-380791 \cos l - 67379 \cos (2D-l).
\end{equation}
Multiplying the $\mathrm{d}A/\mathrm{d}e$ by the time of the observation since the epoch,
we get the approximate partial of the one-way laser range w.r.t. extra $\mathrm{d}e/\mathrm{d}t$.

\subsubsection{Biases}\label{biases-sec}
Biases are determined parameters intended to compensate changes in
station's equipment or other anomalies. Table \ref{tbl-biases} lists
biases used in this work; this set is close to the one used during
building the DE430 ephemerides. Different biases have different origins.

Biases 14, 15, and 16 are known from the changes in calibration
and ranging rings at the Haleakala station (see Table \ref{tbl-haleakala-biases};
while the last two changes did not create any detectable bias). Similarly,
biases 6, 7, and 12 match upgrades of the laser at the CERGA station
(Ruby$\rightarrow$YAG$\rightarrow$MeO). Biases 1, 4, and 5 come from known
changes at the Apache Point station (the installation of a new detector in November 2010;
different calibration technique since April 2012). Biases 13, 21, 27 and 28 cover the
whole timespan of their stations' operation. Other biases have no known cause
(one can guess a human error) and were detected in post-fit residuals.

\begin{table}[h]
  \begin{tabular}{ | r | l | c | c | r | l | c | c | }
    \hline
    \textbf{\#} & \textbf{Station} & \textbf{from}  &  \textbf{to} &
    \textbf{\#} & \textbf{Station} & \textbf{from}  &  \textbf{to}
    \\ \hline
 1 & Apache    & 07.04.2006 & 01.11.2010 & 15 & Haleakala & 02.04.1986 & 30.07.1987 \\ \hline
 2 & Apache    & 15.12.2007 & 30.06.2008 & 16 & Haleakala & 31.07.1987 & 14.08.1987 \\ \hline
 3 & Apache    & 20.09.2008 & 20.06.2009 & 17 & Haleakala & 09.06.1985 & 10.06.1985 \\ \hline
 4 & Apache    & 01.11.2010 & 07.04.2012 & 18 & Haleakala & 28.01.1989 & 29.01.1989 \\ \hline
 5 & Apache    & 07.04.2012 & 02.09.2013 & 19 & Haleakala & 23.08.1989 & 24.08.1989 \\ \hline
 6 & CERGA     & 01.06.1984 & 13.06.1986 & 20 & Haleakala & 06.02.1990 & 01.09.1990 \\ \hline
 7 & CERGA     & 01.10.1987 & 01.08.2005 & 21 & McDonald  & 01.01.1969 & 01.07.1985 \\ \hline
 8 & CERGA     & 10.12.1996 & 18.01.1997 & 22 & McDonald  & 01.12.1971 & 05.12.1972 \\ \hline
 9 & CERGA     & 08.02.1997 & 24.06.1998 & 23 & McDonald  & 21.04.1972 & 27.04.1972 \\ \hline
10 & CERGA     & 04.12.2004 & 07.12.2004 & 24 & McDonald  & 18.08.1974 & 16.10.1974 \\ \hline
11 & CERGA     & 03.01.2005 & 06.01.2005 & 25 & McDonald  & 05.10.1975 & 01.03.1976 \\ \hline
12 & CERGA     & 01.11.2009 & 01.01.2014 & 26 & McDonald  & 01.12.1983 & 17.01.1984 \\ \hline
13 & Haleakala & 01.11.1984 & 01.09.1990 & 27 & Matera    & 01.01.2003 & 01.01.2016 \\ \hline
14 & Haleakala & 01.11.1984 & 01.04.1986 & 28 & MLRS1     & 01.08.1983 & 28.01.1988 \\ \hline
  \end{tabular}
  \caption{Biases determined from LLR observations. Date notation is DD.MM.YYYY}
  \label{tbl-biases}
\end{table}

\begin{table}[h]
  \begin{tabular}{ | c | l | l | }
    \hline
    \textbf{Time span} & \textbf{Calibration rings}  &  \textbf{Ranging rings} \\ \hline
    04.10.1984 -- 01.04.1986 &  2, 3, 4    &    1, 2, 3, 4  \\ \hline
    02.04.1986 -- 30.07.1987 &  3          &    1, 2, 3, 4  \\ \hline
    31.07.1987 -- 14.08.1987 &  3          &          3, 4  \\ \hline
    15.08.1987 -- 09.11.1987 &  3          &          3     \\ \hline
    10.11.1987 -- 18.02.1988 &  3, 4       &          3, 4  \\ \hline
    19.02.1988 -- 31.08.1990 &  1, 2, 3, 4 &    1, 2, 3, 4 \\ \hline
  \end{tabular}
  \caption{Changes of equipment at the Haleakala station. Date notation is DD.MM.YYYY}
  \label{tbl-haleakala-biases}
\end{table}

\subsection{Full list of parameters}
The full list of parameters, fixed or fit, used in different solutions in this work,
is given in Table~\ref{tbl-parameters}. The chosen epoch for determined initial values
is that of the EPM ephemeris: JD~2446000.5, except for stations' positions: their
epochs were chosen individually. Table~\ref{tbl-stations} summarizes the stations'
parameters.

\newcolumntype{C}[1]{>{\centering}p{#1}}

\begin{table}[!htbp]
  \begin{tabularx}{\textwidth}{| C{1.5cm} | X | c | X |} 
    \hline
    \textbf{Notation} & \multicolumn{1}{c|}{\textbf{parameter}} & \textbf{type}  &
                        \multicolumn{1}{c|}{\textbf{notes}}
    \\ \hline
    $\mu_S$ & standard gravitational parameter of the Sun & fixed &
    fixed to DE430 value in this work; may differ in the EPM ephemeris \\ \hline
    $\mu_E/\mu_M$ & Earth-Moon mass ratio & fixed &
    determined from spacecraft observations;
    fixed to DE430 value in this work, may differ in EPM  \\ \hline
    $\mu_E + \mu_M$ & standard gravitational parameter of the E-M system & fit & \\ \hline
    ${\bar C_\mathrm{nm,E}}$, ${\bar S_\mathrm{nm,E}}$ 
    & spherical harmonic coefficients of Earth's gravitational potential & fixed
    & up to $\Nmax = 6$, taken from model based on EGM2008, see section 6.1 of Conventions;
    DE tidal model comes with an altered ${\bar C_\mathrm{20,E}}$  \\ \hline
     $k_{20}$, $k_{21}$, $k_{22}$
     & potential degree-2 Love numbers of Earth zonal, diurnal, and semi-diurnal tides
     & fixed & in DE tidal model: $k_{20}=0.335$, $k_{21}=0.320$, $k_{22}=0.282$; IERS tidal model
    is more complex \\ \hline
    
     $\tau_\mathrm{0O}$, $\tau_\mathrm{1O}$, $\tau_\mathrm{2O}$ 
     & orbital delays of Earth zonal, diurnal, and semi-diurnal tides & fixed/absent
     & only in DE tidal model: $\tau_\mathrm{0O}=0.0780$~d, $\tau_\mathrm{1O}=-0.044$~d,
     $\tau_\mathrm{2O}=-0.113$~d \\ \hline
     
     $\tau_\mathrm{1R}$, $\tau_\mathrm{2R}$ & rotational delays of Earth diurnal, semi-diurnal tides
     & fit/absent & present only in DE tidal model\\ \hline
    $l_2, k_2$ & degree-2 lunar Shida number and Love number & fixed & taken from GRAIL results \\ \hline
    $h_2$ & degree-2 lunar radial displacement Love number & fit &  \\ \hline
    $\bar C_{20}$ & undistorted normalized main zonal lunar harmonic
    & fixed & taken from GRAIL (solution GL660b) \\ \hline
    $\beta, \gamma$ & ratios between undistorted main moments of itertia & fit & \\ \hline
    $\bar C_{21}$, $\bar S_{21}$, $\bar S_{22}$ & other degree-2 harmonics
    & fixed & Zero; $\bar S_{21}$ taken from GL660b in one solution \\ \hline
    $\bar C_{32}$, $\bar S_{32}$, $\bar C_{33}$ & some degree-3 harmonics & fit & \\ \hline
    $\bar C_\mathrm{nm}$, $\bar S_\mathrm{nm}$ & other lunar harmonics  & fixed & taken from GL660b up to degree 6 \\ \hline
    $\tau$ & lunar tidal delay & fit & \\ \hline
    $f_c$ & oblateness of the lunar core & fit & \\ \hline
    $k_v/C_T$ & CMB interaction & fit & \\ \hline
    $\alpha_c$ & core polar moment / undistorted total polar moment & fixed & DE430 fixed value 0.0007 \\ \hline
    $A_1$, $A_2$, $A_3$ & unmodeled longitude libration amplitudes & fit &  \\ \hline
    $\LPA$ ($\times 5$) & positions of five lunar retroreflectors & fit & \\ \hline
    ${\bm r}_\mathrm{EM}$, ${\bm {\dot r}}_\mathrm{EM}$ &
    position and velocity of the Moon w.r.t. Earth in the inertial frame at epoch & fit & \\ \hline
    $\phi$, $\theta$, $\psi$, $\dot\phi$, $\dot\theta$, $\dot\psi$ &
    Euler angles and their rates at epoch  & fit & \\ \hline
    $\STRS$ ($\times 7$), $\bm{\dot s}_\mathrm{TRS}$ ($\times 5$) & positions and velocities of stations at their epochs & fixed/fit
    & see Table~\ref{tbl-stations} \\ \hline
    $\bm{\omega_c}$ & angular velocity of the lunar core at epoch & fit & \\ \hline
    $b$ ($\times 28$) & biases & fit & see Table~\ref{tbl-biases}  \\ \hline
    $\mathrm{d}e/\mathrm{d}t$ & extra eccentricity rate  & fit/absent & present in some solutions \\ \hline
  \end{tabularx}
  \caption{Parameters used in dynamic model or reductions of observations}
  \label{tbl-parameters}
\end{table}

\begin{table}[!htbp]
  \begin{tabularx}{\textwidth}{| X | c | c | X |} 
    \hline
    \multicolumn{1}{|c|}{\textbf{Parameter}} & \textbf{type}  & \multicolumn{1}{c|}{\textbf{notes}} \\ \hline
    McDonald position & fit  & epoch 01.01.1991 \\ \hline
    MLRS1 position & fit  & epoch 01.01.1991 \\ \hline
    MLRS2 position & fit  & epoch 01.01.1991 \\ \hline
    McDonald, MLRS1, MLRS2 velocity & fit & \\ \hline
    Apache position & fit  & epoch 01.06.2009 \\ \hline
    Apache velocity & fixed  & GNSS solution (P027): $(-1.35, 0.03, -0.04)^T$ cm/yr\\ \hline
    CERGA position & fit  & epoch 01.01.2000 \\ \hline
    CERGA velocity & fit  & \\ \hline
    Haleakala position & fit  & epoch 01.04.1986 \\ \hline
    Haleakala velocity & fixed  & GNSS solution: $(-1.30, 6.16, 3.21)^T$ cm/yr\\ \hline
    Matera position & fit  & epoch 01.01.2008 \\ \hline
    Matera velocity & fixed  & GNSS solution: $(-1.85, 1.86, 1.47)^T$ cm/yr\\ \hline
  \end{tabularx}
  \caption{Parameters of stations. Fixed velocities are given in cartesian coordinates}
  \label{tbl-stations}
\end{table}

\section{Software used in this work}
ERA (Ephemeris Research in Astronomy), version 8 was used for processing the observations,
refining the parameters and integrating the dynamical equations \citep{pavlov},
ERA comprises a domain-specific language SLON tailored for astronomical tasks \citep{krasinsky88}.
ERA-8 is a rework of earlier versions of ERA \citep{krasinsky97,krasinsky06}.
Unlike the earlier versions, ERA-8
is based on the Racket programming platform \citep{findler,plt-tr1}.
and has SQLite (\url{http://sqlite.org}) as the database engine.
Most of the numerical algorithms of ERA-8 are implemented in C.

SOFA library (\citealp{hohenkerk}; \url{http://www.iausofa.org}) was used for calculation
of the precession-nutation matrix according
to IAU2000/2006 model, conversion of time scales, calculation of Delaunay arguments,
and conversion between geocentric and geodetic coordinates.

For optical zenith delay \citep{mendes04}
and mapping function \citep{mendes02}, \texttt{FCULZD\_HPA} and \texttt{FCUL\_A} routines
were used. Station displacement due to solid tides \citep{mathews} was calculated with 
the \texttt{DEHANTTIDEINEL} package. For ocean tides, \texttt{HARDISP} package was used;
files with ocean loading coefficients for specific stations were downloaded from the
Onsala Space Observatory website (\url{http://holt.oso.chalmers.se/loading/}).

For numerical integration, an implementation of Gauss-Everhart algorithm from
\citep{avdyushev} was used,
but rewritten from Fortran to C and modified to use extended precision floating-point numbers
(80-bit) instead of double precision (64-bit).

\section{Results}
\subsection{Description of obtained solutions}
Six solutions were obtained in this work. All the solutions are based on the same set of observations,
while differing slightly in dynamical models and determined parameters.

\begin{itemize}
\item Solution I: DE tidal model, $\bar C_{21}^{(0)} = \bar S_{21}^{(0)} = \bar S_{22}^{(0)} = 0$,
  $\mathrm{d}e/\mathrm{d}t$ absent. This model is the closest match to the original DE430 model.
\item Solution II: IERS tidal model, $\bar C_{21}^{(0)} = \bar S_{21}^{(0)} = \bar S_{22}^{(0)} = 0$,
  $\mathrm{d}e/\mathrm{d}t$ absent.
\item Solution III: IERS tidal model, $\bar C_{21}^{(0)}=\bar S_{22}^{(0)}=0$, $\bar S_{21}^{(0)}$
   taken from GL660b, $\mathrm{d}e/\mathrm{d}t$ absent
\item Solution Ie: same as solution I, but with $\mathrm{d}e/\mathrm{d}t$ fit.
\item Solution IIe: same as solution II, but with $\mathrm{d}e/\mathrm{d}t$ fit.
\item Solution IIIe: same as solution III, but with $\mathrm{d}e/\mathrm{d}t$ fit.
  \end{itemize}

The main purpose of obtaining those solutions was to compare IERS and DE tidal models
in terms of their fit to the LLR observations, and to see how they affect the extra
eccentricity rate. Also, it was important to check whether the GL660b mean value of
$\bar S_{21}$ improves the overall results of LLR fits.

\subsection{Difference in accelerations given by IERS and DE tidal models}
Figure~\ref{tidal-forces-de} shows periodic accelerations experienced by the Moon's orbit in Solution I
due to tidal perturbations from Earth. On the same orbit, the tidal accelerations obtained
with the IERS model were calculated. The difference between DE and IERS accelerations (in the lunar frame) is shown
on Figure~\ref{tidal-forces-diff}. DE acceleration is permanently bigger on axis X (towards Earth)
by some 0.02-0.03 $\mathrm{mm}/\mathrm{day}^2$.
This may be due to the K1 tide that has a smaller Love number than the average $k_{21}$
value used for DE430. 

\begin{figure}[!htbp]
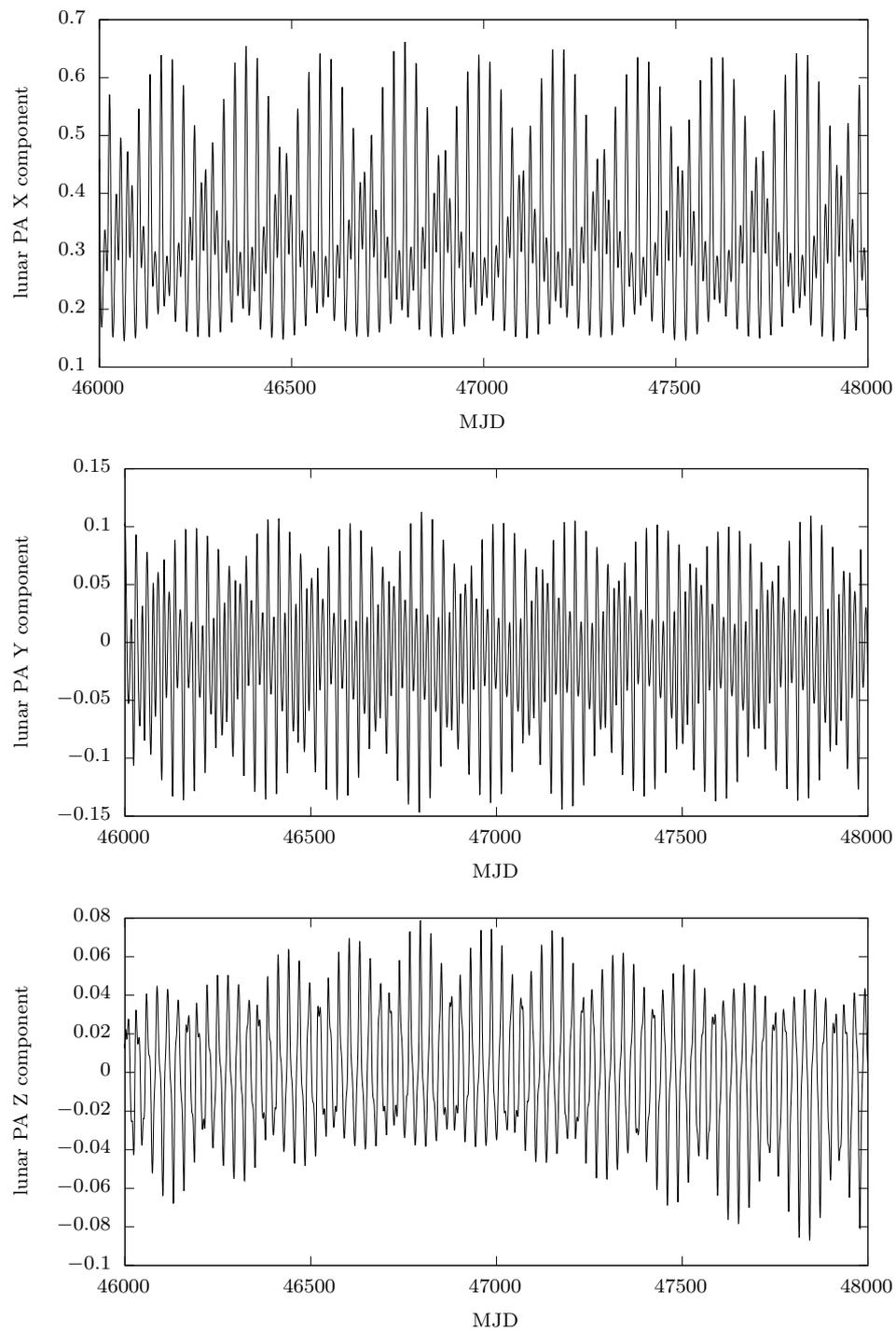

  \caption{Tidal acceleration according to the DE model, lunar frame, $\mathrm{mm}/\mathrm{day}^2$}
  \input tidal_de_x.tex
  
  \input tidal_de_y.tex
  
  \input tidal_de_z.tex
  \label{tidal-forces-de}
  \end{figure}

\begin{figure}
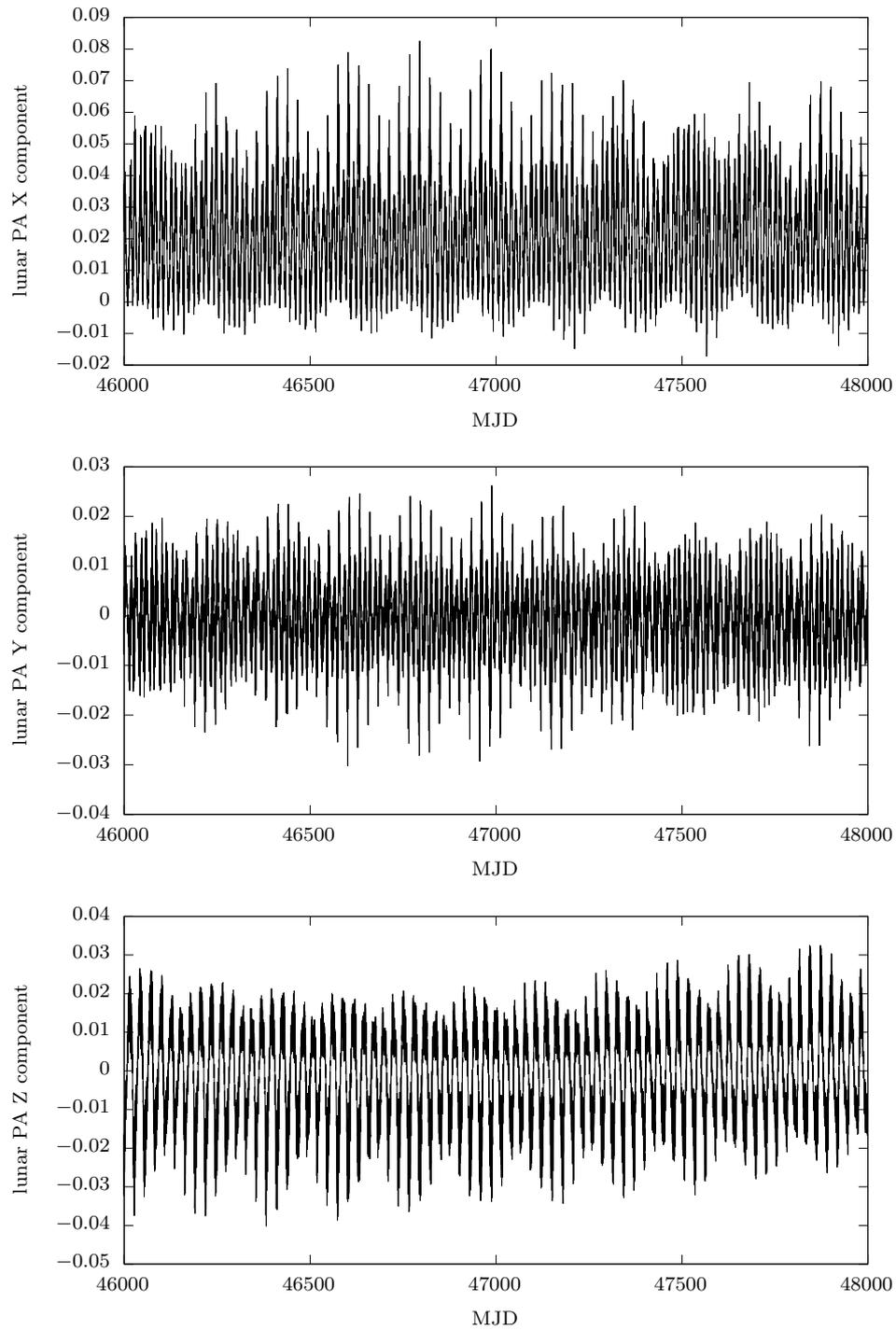

  \caption{DE tidal acceleration minus IERS acceleration on the same orbit,
    lunar frame, $\mathrm{mm}/\mathrm{day}^2$}
  \input tidal_de_iers_diff_x.tex
  
  \input tidal_de_iers_diff_y.tex
  
  \input tidal_de_iers_diff_z.tex
  \label{tidal-forces-diff}
  \end{figure}

\subsection{Post-fit residuals of LLR observations}\label{o-c}
Post-fit statistics of observations for solutions I, II, and III are
shown in Table \ref{tbl-residuals}.  For each station, the number of
utilized normal points is shown, followed by the number of points that have
been automatically rejected, and then by the wrms deviation of $O-C$.

\begin{table}[!htbp]
  \begin{tabular}{ | l | r | r | r | r | r | r |r | r | r |}
    \hline
    & \multicolumn{3}{c|}{\textbf{Solution I}} &
      \multicolumn{3}{c|}{\textbf{Solution II}} &
      \multicolumn{3}{c|}{\textbf{Solution III}}\\ \hline
    \textbf{Station} & used & rej. & wrms & used & rej. & wrms  & used & rej. & wrms \\ \hline
    McDonald     & 3545 & 59  & 19.9 & 3545 & 59  & 20.1 & 3545 & 59  & 20.2 \\ 
    MLRS1        & 587  & 44  & 11.0 & 588  & 43  & 11.3 & 588  & 43  & 11.3 \\ 
    MLRS2        & 3210 & 443 & 3.5  & 3206 & 447 & 3.8  & 3207 & 446 & 3.8 \\ 
    Haleakala    & 748  & 22  & 5.4  & 750  & 20  & 5.8  & 750  & 20  & 5.8 \\ 
    Cerga (Ruby) & 1109 & 79  & 17.2 & 1109 & 79  & 17.5 & 1109 & 79  & 17.5 \\
    Cerga (YAG)  & 8272 & 52  & 2.3  & 8271 & 53  & 2.4  & 8271 & 53  & 2.4 \\
    Cerga (MeO)  & 645  &  9  & 2.2  & 645  & 9   & 2.7  & 645  & 9   & 2.7 \\ 
    Apache       & 1546 & 27  & 1.4  & 1549 & 24  & 1.5  & 1539 & 34  & 1.5 \\
    Matera       & 64   & 19  & 3.8  & 63   & 20  & 3.3  & 63   & 20  & 3.3 \\ \hline
  \end{tabular}
  \caption{Post-fit statistics of solutions I-III. WRMS is one-way and given in cm.} 
  \label{tbl-residuals}
\end{table}
The post-fit statistics of ``e'' counterparts of solutions I-III have
been calculated and found to be nearly the same to as shown in
Table~\ref{tbl-residuals}, and are not listed here.

Plots of one-way $O-C$ of processed observations for selected stations are presented:
Figure~\ref{o-c-mcdonald} for McDonald, MLRS1, and MLRS2,
Figure~\ref{o-c-cerga} for CERGA, and Figure~\ref{o-c-apache} for
Apache Point.

\begin{figure}[!htbp]
  \caption{Post-fit residuals for McDonald, MLRS1, and MLRS2 stations in Solution I}
  \input residuals-mcdonald.tex
  \label{o-c-mcdonald}
  \end{figure}

\begin{figure}[!htbp]
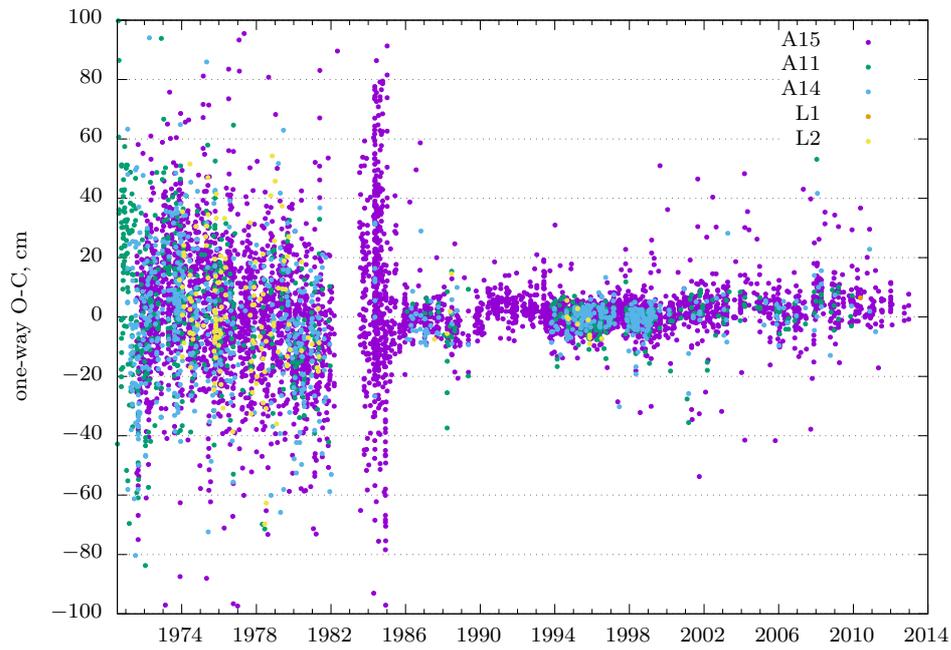

  \caption{Post-fit residuals for CERGA station in Solution I}
  \input residuals-cerga.tex
  \label{o-c-cerga}
  \end{figure}

\begin{figure}[!htbp]
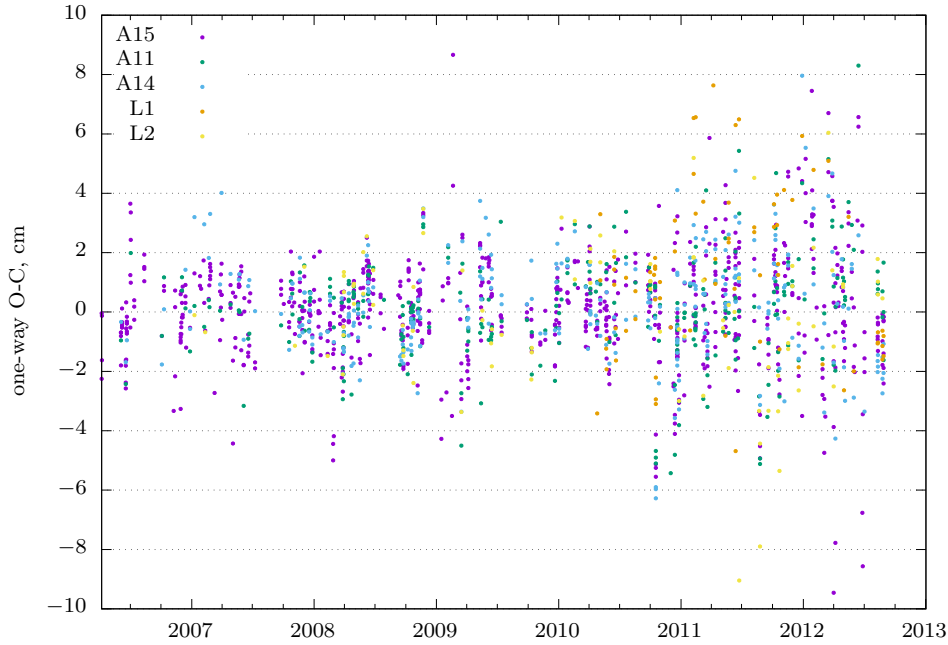

  \caption{Post-fit residuals for Apache Point station in Solution I}
  \input residuals-apache.tex
  \label{o-c-apache}
  \end{figure}

\subsection{Determined parameters}
Parameters determined in solutions I-III, along with their formal uncertainties,
are listed in Tables~\ref{tbl-determined-1},~\ref{tbl-determined-2}, and~\ref{tbl-determined-3}.
The initial parameters (except for stations' positions) are given at epoch JD~2446000.5.

\begin{table}[h]
  \setlength{\tabcolsep}{2.5pt}
  \begin{tabular}{ | l | r | r | r | l | }
    \hline
    \textbf{Parm.} & \multicolumn{1}{c|}{\textbf{Solution I value}} &
    \multicolumn{1}{c|}{\textbf{Solution II value}} &
     \multicolumn{1}{c|}{\textbf{Solution III value}}  & \textbf{units} \\ \hline
    ${{\bm r}_\mathrm{EM}}$.x & $-137136474.05\pm0.05$ & $-137136473.33\pm0.05$ & $-137136473.43\pm0.06$ & m\\
    ${{\bm r}_\mathrm{EM}}$.y & $-311514604.01\pm0.05$ & $-311514604.25\pm0.05$ & $-311514604.22\pm0.06$ & m\\
    ${{\bm r}_\mathrm{EM}}$.z & $-141738600.43\pm0.04$ & $-141738600.26\pm0.05$ & $-141738600.17\pm0.05$ & m\\
     ${\bm{\dot r}}_\mathrm{EM}$.x & $962372276.11\pm0.13$   & $962372276.35\pm0.14$ & $962372276.14\pm0.14$ &
     $\mu$m/sec\\ 
     ${\bm{\dot r}}_\mathrm{EM}$.y & $-375608190.19\pm0.14$  & $-375608188.59\pm0.15$ & $-375608188.88\pm0.14$ &
     $\mu$m/sec\\
     ${\bm{\dot r}}_\mathrm{EM}$.z & $-268439311.42\pm0.06$  & $-268439310.06\pm0.06$ & $-268439310.14\pm0.07$ &
     $\mu$m/sec\\ \hline
     $\bm{\omega_c}$.x & $(-890\pm4)\cdot10^{-6}$  &
                             $(-920\pm4)\cdot 10^{-6}$ &
                             $(-932\pm4)\cdot 10^{-6}$ & rad/day\\
     $\bm{\omega_c}$.y & $(-6453\pm8)\cdot10^{-6}$  &
                             $(-6496\pm8)\cdot 10^{-6}$ &
                             $(-6484\pm8)\cdot 10^{-6}$& rad/day\\
     $\bm{\omega_c}$.z & $(229.63\pm0.05)\cdot10^{-3}$ &
                             $(230.32\pm0.03)\cdot10^{-3}$ &
                             $(230.22\pm0.02)\cdot10^{-3}$ & rad/day\\ \hline
     $\phi$ & $(-5823800\pm2)\cdot 10^{-8}$
            & $(-5823802\pm2)\cdot 10^{-8}$
            & $(-5823821\pm2)\cdot 10^{-8}$ & rad \\
     $\theta$ & $(39511625\pm1)\cdot10^{-8}$ & $(39511623\pm1)\cdot 10^{-8}$& $(39511618\pm 1)\cdot 10^{-8}$& rad \\ 
     $\psi$ & $(113574562\pm3)\cdot10^{-8}$ &
              $(113574584\pm3)\cdot10^{-8}$&
              $(113574591\pm3)\cdot10^{-8}$ & rad \\ 
     $\dot\phi$ & $-74.537\pm0.001$ & $-74.541\pm0.001$ & $-74.543\pm0.001$ & $\arcsec$/day \\
     $\dot\theta$ & $-37.0255\pm0.0003$ & $-37.0257\pm0.0004$ & $-37.0227\pm0.0004$ & $\arcsec$/day \\
     $\dot\psi$ & $47501.853\pm0.001$ & $47501.859\pm0.001$ & $47501.862\pm0.001$ & $\arcsec$/day \\ \hline
     $\mu_E+\mu_M$ & $403503.2365\pm0.0002$& $403503.2360\pm0.0002$ &
     $403503.2358\pm0.0002$ & $\mathrm{km}^3/\mathrm{s}^2$ \\ \hline
     $\beta$ & $(631023.1 \pm 0.5)\cdot10^{-9}$ & $(631024.5\pm0.5)\cdot10^{-9}$& $(631024.9\pm0.5)\cdot10^{-9}$ & 1 \\ 
     $\gamma$ & $(227733.3 \pm 0.7)\cdot10^{-9}$ &
                $(227736.0\pm0.7)\cdot10^{-9}$ &
                $(227736.3\pm0.7)\cdot10^{-9}$ & 1 \\ \hline 
     $\tau$ & $0.096\pm0.001$ & $0.079\pm0.001$ & $0.075\pm0.001$ & day \\ \hline
     $\tau_\mathrm{1R}$ & $0.00787\pm0.00005$ & N/A & N/A & day \\
     $\tau_\mathrm{2R}$ & $0.002855\pm0.000004$ & N/A &  N/A & day \\ \hline
     $f_c$ & $(0.247\pm0.004)\cdot10^{-3}$ &
             $(0.249\pm0.004)\cdot10^{-3}$ &
             $(0.245\pm0.004)\cdot10^{-3}$ & 1 \\ 
     $k_v/C_T$ & $(16.3 \pm 0.2)\cdot10^{-9}$ &
                $(18.6\pm0.2)\cdot10^{-9}$ &
                $(19.4\pm0.2)\cdot10^{-9}$ & $\textrm{day}^{-1}$ \\ \hline
     $h_2$ & $0.043\pm 0.001$ & $0.041\pm 0.001$ & $0.041\pm 0.001$  & 1 \\ \hline
     $A_1$ & $4.6\pm0.2$ &$4.3\pm0.2$  & $4.4\pm0.2$& mas \\
     $A_2$ & $1.4\pm0.2$ &$0.7\pm0.2$ & $0.4\pm0.2$ & mas \\
     $A_3$ & $-7.4\pm0.5$ & $-10.3\pm0.5$& $-12.0\pm0.5$& mas \\ \hline
     ${\bar C}_{32}$ & $(14184.3\pm0.3)\cdot10^{-9}$ &
                      $(14185.5\pm0.4)\cdot10^{-9}$ &
                      $(14184.9\pm0.3)\cdot10^{-9}$ & 1 \\
     ${\bar S}_{32}$ & $(4931.8\pm0.6)\cdot10^{-9}$ &
                      $(4937.4\pm0.7)\cdot10^{-9}$ &
                      $(4896.3\pm0.6)\cdot10^{-9}$ & 1 \\ 
     ${\bar C}_{33}$ & $(11975\pm11)\cdot10^{-9}$ &
                      $(11912\pm11)\cdot10^{-9}$ &
                      $(11913\pm11)\cdot10^{-9}$
     & 1 \\ \hline
     A11 x & $1591966.90\pm0.06$ & $1591966.80\pm0.06$ & $1591966.72\pm0.06$ & m \\ 
     A11 y & $ 690699.56\pm0.04$ & $690699.32\pm0.04$ & $690699.47\pm0.04$ & m \\
     A11 z & $  21003.73\pm0.02$ & $21003.76\pm0.02$ & $21003.78\pm0.02$& m \\ \hline
     A14 x & $1652689.86\pm0.06$ & $1652689.63\pm0.07$ & $1652689.65\pm0.06$ & m \\
     A14 y & $-520997.46\pm0.04$ & $-520997.78\pm0.05$ & $-520997.58\pm0.04$ & m \\
     A14 z & $-109730.52\pm0.02$ & $-109730.47\pm0.02$ & $-109730.52\pm0.02$ & m \\ \hline
     A15 x & $1554678.60\pm0.07$ & $1554678.41\pm0.07$ & $1554678.39\pm0.06$ & m \\
     A15 y & $  98095.64\pm0.04$ & $98095.32\pm0.04$ & $98095.26\pm0.04$ & m \\ 
     A15 z & $ 765005.14\pm0.03$ & $765005.19\pm0.03$ & $765005.17\pm0.04$ & m \\ \hline
     L1 x &  $1114292.57\pm0.06$ & $1114292.29\pm0.06$ & $1114292.35\pm0.06$ & m \\
     L1 y & $ -781298.47\pm0.04$ & $-781298.82\pm0.04$ & $-781298.99\pm0.04$ & m \\
     L1 z & $ 1076058.49\pm0.03$ & $1076058.38\pm0.04$ & $1076058.34\pm0.06$ & m \\ \hline
     L2 x & $ 1339363.66\pm0.06$ & $1339363.56\pm0.06$ & $1339363.48\pm0.06$ & m \\
     L2 y & $  801872.04\pm0.04$ & $801871.80\pm0.04$ & $801871.73\pm0.04$& m \\ 
     L2 z &  $ 756358.60\pm0.03$ & $756358.62\pm0.03$ & $756358.65\pm0.03$ & m \\ \hline
 \end{tabular}
  \caption{Determined parameters and their formal uncertainties, part 1}
  \label{tbl-determined-1}
\end{table}

\begin{table}[h]
  \setlength{\tabcolsep}{2.5pt}
  \begin{tabular}{ | l | r | r | r | l | }
    \hline
    \textbf{Parameter} & \multicolumn{1}{c|}{\textbf{Solution I value}} &
    \multicolumn{1}{c|}{\textbf{Solution II value}} &
    \multicolumn{1}{c|}{\textbf{Solution III value}}  & \textbf{units} \\ \hline
    
     McD $\lambda$ &  $255.978002(5\pm1)$ & $255.978002(3\pm1)$ & $255.978002(3\pm1)$& $\Deg$ \\
     McD $r\cos\phi$&  $5492414.46\pm0.03$   & $5492414.45\pm0.03$ & $5492414.45\pm0.03$ & m \\ 
     McD $r\sin\phi$& $3235697.50\pm0.02$    & $3235697.50\pm0.02$ & $3235697.51\pm0.02$ & m \\ \hline
     
     MLRS1 $\lambda$& $255.984120(8\pm1)$ & $255.984120(9\pm1)$ & $255.984120(9\pm1)$ & $\Deg$ \\
     MLRS1 $r\cos\phi$& $5492037.71\pm0.03$  & $5492037.67\pm0.04$ & $5492037.67\pm0.04$ & m \\
     MLRS1 $r\sin\phi$& $3236146.76\pm0.02$  & $3236146.75\pm0.02$ & $3236146.75\pm0.02$ & m \\ \hline
     
     MLRS2 $\lambda$& $255.9848036(6\pm3)$& $255.9848036(5\pm3)$ & $255.9848036(5\pm3)$ & $\Deg$ \\
     MLRS2 $r\cos\phi$&$5491888.44\pm0.01$ & $5491888.44\pm0.01$ & $5491888.43\pm0.01$ & m \\ 
     MLRS2 $r\sin\phi$&$3236481.64\pm0.01$ & $3236481.62\pm0.01$ & $3236481.63\pm0.01$ & m \\ \hline

     Apache $\lambda$& $254.17957680(8\pm7)$ & $254.17957679(3\pm8)$ & $254.17957679(2\pm8)$ & $\Deg$ \\
     Apache $r\cos\phi$ &$5370045.373\pm0.002$ & $5370045.376\pm0.002$ & $5370045.378\pm0.002$ & m \\ 
     Apache $r\sin\phi$ &$3435012.897\pm0.002$ & $3435012.913\pm0.002$ & $3435012.910\pm0.002$ & m \\ \hline

     CERGA $\lambda$& $6.9215727(8\pm1)$ & $6.9215727(5\pm1)$ & $6.9215727(5\pm1)$ & $\Deg$ \\ 
     CERGA $r\cos\phi$& $4615328.454\pm0.002$ & $4615328.450\pm0.002$ & $4615328.450\pm0.002$ & m \\
     CERGA $r\sin\phi$& $4389355.103\pm0.003$ & $4389355.106\pm0.003$ & $4389355.107\pm0.003$ & m \\ \hline

     Haleakala $\lambda$& $203.7440954(3\pm3)$ & $203.7440955(6\pm3)$ & $203.7440955(8\pm3)$ & $\Deg$ \\
     Haleakala $r\cos\phi$& $5971474.51\pm0.01$ & $5971474.53\pm0.01$ & $5971474.53\pm0.01$ & m \\
     Haleakala $r\sin\phi$& $2242188.41\pm0.01$ & $2242188.43\pm0.01$ & $2242188.44\pm0.01$ & m \\ \hline

     Matera $\lambda$& $16.704613(5\pm7)$ & $16.704613(3\pm2)$ & $16.704613(3\pm2)$ & $\Deg$ \\
     Matera $r\cos\phi$& $4846504.3\pm0.2$ & $4846504.25\pm0.04$ & $4846504.24\pm0.04 $ & m \\
     Matera $r\sin\phi$& $4133249.58\pm0.05$ & $4133249.59\pm0.01$ & $4133249.58\pm0.02$ & m \\ \hline

     McD $\dot\lambda$ &  $-0.53 \pm 0.01$ & $-0.56\pm0.01$ & $-0.56\pm0.01$ & mas/yr \\
     McD $(r\cos\phi)\dot{}$ &  $3.5\pm0.2$ & $2.8\pm0.2$ & $2.9\pm0.2$ & mm/yr \\ 
     McD $(r\sin\phi)\dot{}$ &  $3.5\pm0.5$ & $4.2\pm0.5$ & $3.7\pm0.5$ & mm/yr \\ \hline
     CERGA $\dot\lambda$ &  $0.915\pm 0.007$ & $0.914\pm0.007$ & $0.913\pm0.008$ & mas/yr \\ 
     CERGA $(r\cos\phi)\dot{}$ &  $-15.7\pm0.2$ & $-16.5\pm0.2$ & $-16.5\pm0.2$ & mm/yr \\ 
     CERGA $(r\sin\phi)\dot{}$ &  $14.3\pm0.4$ & $13.9\pm0.4$ & $13.8\pm0.4$ & mm/yr \\ \hline
\end{tabular}
  \caption{Determined parameters and their formal uncertainties, part 2: stations}
  \label{tbl-determined-2}
\end{table}

\begin{table}[h]
  \setlength{\tabcolsep}{2.5pt}
  \begin{tabular}{ | l | r | r | r | l | }
    \hline
    \textbf{Parameter} & \multicolumn{1}{c|}{\textbf{Solution I value}} &
    \multicolumn{1}{c|}{\textbf{Solution II value}} &
    \multicolumn{1}{c|}{\textbf{Solution III value}}  & \textbf{units} \\ \hline
    
     Bias 1 (APOLLO) &   $6.4\pm1.1$ &   $7.7\pm1.2$ & $8.2\pm1.2$ & cm \\ \hline
     Bias 2 (APOLLO) &   $10.9\pm1.1$ &  $12.6\pm1.2$ & $13.1\pm1.2$ & cm \\ \hline
     Bias 3 (APOLLO) &   $3.3\pm1.1$ &   $5.0\pm1.2$ & $5.3\pm1.2$ & cm \\ \hline
     Bias 4 (APOLLO) &   $12.1\pm1.2$ &  $12.6\pm1.3$ & $13.1\pm1.3$ & cm \\ \hline
     Bias 5 (APOLLO) &   $-1.7\pm1.1$ &  $-1.1\pm1.2$ & $-0.4\pm1.2$ & cm \\ \hline
     Bias 6 (CERGA) &   $23.1\pm1.9$ &  $30.8\pm2.1$ & $30.6\pm2.0$ & cm \\ \hline
     Bias 7 (CERGA) &   $4.1\pm1.1$ &   $5.2\pm1.2$ & $5.2\pm1.2$ & cm \\ \hline
     Bias 8 (CERGA) &   $-11.7\pm1.5$ &  $-7.3\pm1.6$ & $-7.2\pm1.6$ & cm \\ \hline
     Bias 9 (CERGA) &   $-13.5\pm1.2$ & $-11.5\pm1.2$ & $-11.7\pm1.2$ & cm \\ \hline
     Bias 10 (CERGA) &   $6.0\pm1.7$ &  $11.8\pm1.8$ & $11.2\pm1.8$ & cm \\ \hline
     Bias 11 (CERGA) &   $4.8\pm2.1$ &  $11.7\pm2.2$ & $11.7\pm2.2$ & cm \\ \hline
     Bias 12 (CERGA) &   $-4.0\pm1.2$ &  $-6.0\pm1.3$ & $-5.5\pm1.3$ & cm \\ \hline
     Bias 13 (Haleakala) &   $-0.6\pm1.7$ &   $1.9\pm1.8$ & $2.1\pm1.7$ & cm \\ \hline
     Bias 14 (Haleakala)  &  $8.3\pm1.9$ &  $14.3\pm0.2$ & $14.6\pm2.0$ & cm \\ \hline
     Bias 15 (Haleakala)  &  $-11.7\pm1.8$ &  $-9.1\pm1.9$ & $-9.0\pm1.9$ & cm \\ \hline
     Bias 16 (Haleakala)  &  $3.0\pm2.4$ &   $2.9\pm2.5$ & $3.3\pm2.5$ & cm \\ \hline
     Bias 17 (Haleakala)  &  $38.3\pm4.0$ &  $37.7\pm4.3$ & $37.8\pm4.2$ & cm \\ \hline
     Bias 18 (Haleakala)  &  $40.5\pm3.2$ &  $47.1\pm3.4$ & $48.2\pm3.3$ & cm \\ \hline
     Bias 19 (Haleakala)  &  $24.3\pm1.9$ &  $27.4\pm2.1$ & $27.5\pm2.0$ & cm \\ \hline
     Bias 20 (Haleakala)  &  $-9.5\pm1.8$ &  $-3.8\pm1.9$ & $-3.4\pm1.9$ & cm \\ \hline
     Bias 21 (McDonald)  &   $42.0\pm4.9$ &  $42.3\pm5.2$ & $42.6\pm5.2$ & cm \\ \hline
     Bias 22 (McDonald)  &   $28.3\pm5.3$ &  $29.5\pm5.7$ & $30.0\pm5.7$ & cm \\ \hline
     Bias 23 (McDonald)  &  $-52.8\pm18.3$ & $-52.5\pm19.6$ & $-53.2\pm19.4$ & cm \\ \hline
     Bias 24 (McDonald)  &  $160.1\pm6.8$ & $162.6\pm7.3$  & $163.8\pm7.3$ & cm \\ \hline
     Bias 25 (McDonald)  &  $22.2\pm6.3$ &  $20.7\pm6.8$  & $22.0\pm6.7$ & cm \\ \hline
     Bias 26 (McDonald)  &  $-22.4\pm14.9$ & $-11.1\pm16.0$ & $-11.0\pm15.8$ & cm \\ \hline
     Bias 27 (Matera) &   $23.6\pm35.8$ &  $5.1\pm7.6$   & $3.7\pm7.5$ & cm \\ \hline
     Bias 28 (MLRS1)  &   $6.0\pm6.2$ &  $1.8\pm6.6$   & $2.1\pm6.5$ & cm \\ \hline
\end{tabular}
  \caption{Determined parameters and their formal uncertainties, part 3: biases (two-way)}
  \label{tbl-determined-3}
\end{table}

Table~\ref{tbl-eccentricity} shows the extra eccentricity rates found
in three ``e'' solutions.
\begin{table}[!htbp]
  \begin{tabular}{ | c | r | r |}
    \hline
    \textbf{Solution} & \textbf{extra} $\mathrm{d}e/\mathrm{d}t$, $\mathrm{yr}^{-1}$ &
    $1\sigma$, $\mathrm{yr}^{-1}$ \\ \hline
  Ie   & $1.4\times 10^{-12}$  & $0.2\times 10^{-12}$\\ \hline
  IIe  & $-1.3\times 10^{-12}$ & $0.2\times 10^{-12}$\\ \hline
  IIIe & $-1.4\times 10^{-12}$ & $0.2\times 10^{-12}$\\ \hline
  \end{tabular}
  \caption{Extra eccentricity rate and its formal uncertainty in obtained solutions} 
  \label{tbl-eccentricity}
\end{table}

\subsection{Derived parameters}
Secular tidal perturbation terms of the Earth-Moon system have been derived from Solution I
using a table that converts the Love numbers and time delays \citep{williams16},
tidal acceleration $\mathrm{d}n/\mathrm{d}t \approx -25.901\, \arcsec/\mathrm{cy}^2$, semimajor axis
rate $\mathrm{d}a/\mathrm{d}t \approx 38.204\, \mathrm{mm}/\mathrm{yr}$, and the
eccentricity rate $\mathrm{d}e/\mathrm{d}t \approx 13.4\times 10^{-12}\, /\mathrm{yr}$.
The last figure is the modeled eccentricity rate, not including the found extra
$\mathrm{d}e/\mathrm{d}t$.

The lunar $\tau$ from Solution I indicates that $k_2/Q=5.34\times 10^{-4}$ or $Q=45$
at a~1~month period, where $Q$ is the tidal quality factor.
The annual $A_1$ parameter gives $k_2/Q=5.6\times 10^{-4}$ or $Q=45$ at a~1~year period.

\medskip

\noindent The mean $\bar C_{22}$ value can be calculated from determined $\beta$ and $\gamma$:
\begin{equation}
  \bar C_{22} = -\bar C_{20}\frac{\gamma (1+\beta)}{2(2\beta-\gamma+\beta\gamma)}\frac{N_{20}}{N_{22}}.
  \end{equation}
Taking $\beta$ and $\gamma$ from solution I, one can find $\bar C_{22} = 0.346754\times10^{-4}$.

\medskip

\noindent From the $\mu_E+\mu_M$ found in Solution I, and $\mu_E/\mu_M$ fixed to
81.30056907, one can calculate $\mu_M = 4902.80008\ \mathrm{km}^3/\mathrm{s}^2$ and
$\mu_E = 398600.4364\ \mathrm{km}^3/\mathrm{s}^2$.

\section{Conclusion}

The results of this work can be summarized as follows:
\begin{itemize}
\item Full implementation of DE430 lunar model was obtained and built into the EPM ephemeris
  software;
\item The conventional model of Earth's gravitational potential has proven 
  suitable for analyzing LLR observations;
\item The IAU2000/2006 precession-nutation model along with the IERS C04 EOP series
  have proven suitable for analyzing LLR observations, with the only exception
  of IERS C04 before 1984 (JPL KEOF EOP series may be used instead);
\item Secular tidal perturbation terms of the Earth-Moon system have been calculated:
  $\mathrm{d}n/\mathrm{d}t \approx -25.901\, \arcsec/\mathrm{cy}^2$, $\mathrm{d}a/\mathrm{d}t \approx 38.204\, \mathrm{mm}/\mathrm{yr}$,
  modeled $\mathrm{d}e/\mathrm{d}t \approx 13.4\times 10^{-12}\, /\mathrm{yr}$,
  extra $\mathrm{d}e/\mathrm{d}t \approx 1.4\times 10^{-12}\, /\mathrm{yr}$.

\item Models of tidal station displacement from solid Earth tides
  and tropospheric delay, recommended in the IERS
  Conventions 2010, has proven suitable for analyzing LLR observations.
  The ocean loading model, though put to use, has not been checked thoroughly;
  atmospheric loading and ocean pole tide loading were not implemented;
\item In addition to the DE430 model of tidal acceleration of the orbit of the Moon,
  the ``IERS 2010'' model of tidal variations of the geopotential has been
  implemented. It has been found that the IERS model fits slightly worse
  to LLR observations, though that was not unexpected given
  two fewer solution parameters. It has been found that lunar $\tau$ and extra
  $\mathrm{d}e/\mathrm{d}t$ are very sensitive to the tidal model used;
  in particular, the extra eccentricity rate falls from $1.4\times10^{-12}$
  with the DE model to $-1.3\times 10^{-12}$ with the IERS model;
\item Non-zero mean value of lunar $\bar S_{21}^{(0)}$, taken from GRAIL,
  does not make any significant difference in postfit results and
  does not affect much the determined parameters of lunar inner structure;
  similar tests (not shown in the results) were done for
  $\bar C_{21}^{(0)}$ and $\bar S_{22}^{(0)}$, with similar outcomes.
\item The strong detection of $k_v/C_T$ demonstrates that the Moon has a fluid core.
\item The lunar $\tau$ indicates substantial tidal dissipation with $Q=45$ at a 1-month period
  and the annual $A_1$ parameter shows similarly strong dissipation with $Q=45$ at a 1-year period.
\item Determined $\bar C_{32}$ differs from the GL660b value
  ($1.41715\times10^{-5}$) by $<0.1\%$;
  derived $\bar C_{22}$ is also very close to GL660b value ($0.346737\times10^{-4}$);
\item The determined $\bar S_{32}$ value differs from the GL660b
  value ($4.8780\times10^{-6}$) by $0.4$--$1.2\%$, depending on the solution; the determined
  $\bar C_{33}$ value differs by some $3\%$ from the GL660b value ($1.2275\times10^{-5}$).
\end{itemize}

More research is needed to find the cause of the difference between the values
of $\bar S_{32}$ and $\bar C_{33}$ determined from LLR and GRAIL, and
the cause of the misalignment of the lunar PA frame in the model with the GRAIL's frame.
A separate direction of research is the influence of the IERS tidal model on the
eccentricity rate.

\begin{acknowledgements}
D. Pavlov would like to thank Elena Pitjeva, Eleonora Yagudina, Sergey Kurdubov,
Vladimir Skripnichenko, and numerous other colleagues from the IAA RAS for helpful comments
and advice throughout this work; and Matthew Flatt from the University of Utah for his help
in programming on the Racket platform.

This work would not have been possible without the effort of personnel at observatories
doing lunar laser ranging: Apache Point \citep{murphy12,murphy13},
McDonald Laser Ranging Station \citep{shelus},
Observatoire de la C\^ote d'Azur \citep{samain},
Giuseppe Bianco at Matera Laser Ranging Observatory, and Lunar Ranging Experiment (LURE) at the
Haleakala observatory in the past. The POLAC website was of great help, where
Christophe Barache, S\'ebastien Bouquillon, Teddy Carlucci, and Gerard Francou
carefully collected LLR observations from different sources.

An anonymous reviewer provided a lot of comments and suggestions that allowed
to improve the article substantially.

A portion of the research described in this paper was carried out at the 
Jet Propulsion Laboratory of the California Institute of Technology, 
under a contract with the National Aeronautics and Space Administration. 
Government sponsorship acknowledged.
\end{acknowledgements}

\bibliography{pavlov-llr}

\end{document}